\documentclass[usenatbib,usegraphicx]{mn2e}
\usepackage{comment}
\usepackage{amsfonts}
\usepackage{amsmath}
\usepackage{amssymb}
\usepackage{color}
\definecolor{AliceBlue}{rgb}{0.94,0.97,1.00}
\definecolor{AntiqueWhite1}{rgb}{1.00,0.94,0.86}
\definecolor{AntiqueWhite2}{rgb}{0.93,0.87,0.80}
\definecolor{AntiqueWhite3}{rgb}{0.80,0.75,0.69}
\definecolor{AntiqueWhite4}{rgb}{0.55,0.51,0.47}
\definecolor{AntiqueWhite}{rgb}{0.98,0.92,0.84}
\definecolor{BlanchedAlmond}{rgb}{1.00,0.92,0.80}
\definecolor{BlueViolet}{rgb}{0.54,0.17,0.89}
\definecolor{CadetBlue1}{rgb}{0.60,0.96,1.00}
\definecolor{CadetBlue2}{rgb}{0.56,0.90,0.93}
\definecolor{CadetBlue3}{rgb}{0.48,0.77,0.80}
\definecolor{CadetBlue4}{rgb}{0.33,0.53,0.55}
\definecolor{CadetBlue}{rgb}{0.37,0.62,0.63}
\definecolor{CornflowerBlue}{rgb}{0.39,0.58,0.93}
\definecolor{DarkBlue}{rgb}{0.00,0.00,0.55}
\definecolor{DarkCyan}{rgb}{0.00,0.55,0.55}
\definecolor{DarkGoldenrod1}{rgb}{1.00,0.73,0.06}
\definecolor{DarkGoldenrod2}{rgb}{0.93,0.68,0.05}
\definecolor{DarkGoldenrod3}{rgb}{0.80,0.58,0.05}
\definecolor{DarkGoldenrod4}{rgb}{0.55,0.40,0.03}
\definecolor{DarkGoldenrod}{rgb}{0.72,0.53,0.04}
\definecolor{DarkGray}{rgb}{0.66,0.66,0.66}
\definecolor{DarkGreen}{rgb}{0.00,0.39,0.00}
\definecolor{DarkGrey}{rgb}{0.66,0.66,0.66}
\definecolor{DarkKhaki}{rgb}{0.74,0.72,0.42}
\definecolor{DarkMagenta}{rgb}{0.55,0.00,0.55}
\definecolor{DarkOliveGreen1}{rgb}{0.79,1.00,0.44}
\definecolor{DarkOliveGreen2}{rgb}{0.74,0.93,0.41}
\definecolor{DarkOliveGreen3}{rgb}{0.64,0.80,0.35}
\definecolor{DarkOliveGreen4}{rgb}{0.43,0.55,0.24}
\definecolor{DarkOliveGreen}{rgb}{0.33,0.42,0.18}
\definecolor{DarkOrange1}{rgb}{1.00,0.50,0.00}
\definecolor{DarkOrange2}{rgb}{0.93,0.46,0.00}
\definecolor{DarkOrange3}{rgb}{0.80,0.40,0.00}
\definecolor{DarkOrange4}{rgb}{0.55,0.27,0.00}
\definecolor{DarkOrange}{rgb}{1.00,0.55,0.00}
\definecolor{DarkOrchid1}{rgb}{0.75,0.24,1.00}
\definecolor{DarkOrchid2}{rgb}{0.70,0.23,0.93}
\definecolor{DarkOrchid3}{rgb}{0.60,0.20,0.80}
\definecolor{DarkOrchid4}{rgb}{0.41,0.13,0.55}
\definecolor{DarkOrchid}{rgb}{0.60,0.20,0.80}
\definecolor{DarkRed}{rgb}{0.55,0.00,0.00}
\definecolor{DarkSalmon}{rgb}{0.91,0.59,0.48}
\definecolor{DarkSeaGreen1}{rgb}{0.76,1.00,0.76}
\definecolor{DarkSeaGreen2}{rgb}{0.71,0.93,0.71}
\definecolor{DarkSeaGreen3}{rgb}{0.61,0.80,0.61}
\definecolor{DarkSeaGreen4}{rgb}{0.41,0.55,0.41}
\definecolor{DarkSeaGreen}{rgb}{0.56,0.74,0.56}
\definecolor{DarkSlateBlue}{rgb}{0.28,0.24,0.55}
\definecolor{DarkSlateGray1}{rgb}{0.59,1.00,1.00}
\definecolor{DarkSlateGray2}{rgb}{0.55,0.93,0.93}
\definecolor{DarkSlateGray3}{rgb}{0.47,0.80,0.80}
\definecolor{DarkSlateGray4}{rgb}{0.32,0.55,0.55}
\definecolor{DarkSlateGray}{rgb}{0.18,0.31,0.31}
\definecolor{DarkSlateGrey}{rgb}{0.18,0.31,0.31}
\definecolor{DarkTurquoise}{rgb}{0.00,0.81,0.82}
\definecolor{DarkViolet}{rgb}{0.58,0.00,0.83}
\definecolor{DeepPink1}{rgb}{1.00,0.08,0.58}
\definecolor{DeepPink2}{rgb}{0.93,0.07,0.54}
\definecolor{DeepPink3}{rgb}{0.80,0.06,0.46}
\definecolor{DeepPink4}{rgb}{0.55,0.04,0.31}
\definecolor{DeepPink}{rgb}{1.00,0.08,0.58}
\definecolor{DeepSkyBlue1}{rgb}{0.00,0.75,1.00}
\definecolor{DeepSkyBlue2}{rgb}{0.00,0.70,0.93}
\definecolor{DeepSkyBlue3}{rgb}{0.00,0.60,0.80}
\definecolor{DeepSkyBlue4}{rgb}{0.00,0.41,0.55}
\definecolor{DeepSkyBlue}{rgb}{0.00,0.75,1.00}
\definecolor{DimGray}{rgb}{0.41,0.41,0.41}
\definecolor{DimGrey}{rgb}{0.41,0.41,0.41}
\definecolor{DodgerBlue1}{rgb}{0.12,0.56,1.00}
\definecolor{DodgerBlue2}{rgb}{0.11,0.53,0.93}
\definecolor{DodgerBlue3}{rgb}{0.09,0.45,0.80}
\definecolor{DodgerBlue4}{rgb}{0.06,0.31,0.55}
\definecolor{DodgerBlue}{rgb}{0.12,0.56,1.00}
\definecolor{FloralWhite}{rgb}{1.00,0.98,0.94}
\definecolor{ForestGreen}{rgb}{0.13,0.55,0.13}
\definecolor{GhostWhite}{rgb}{0.97,0.97,1.00}
\definecolor{GreenYellow}{rgb}{0.68,1.00,0.18}
\definecolor{HotPink1}{rgb}{1.00,0.43,0.71}
\definecolor{HotPink2}{rgb}{0.93,0.42,0.65}
\definecolor{HotPink3}{rgb}{0.80,0.38,0.56}
\definecolor{HotPink4}{rgb}{0.55,0.23,0.38}
\definecolor{HotPink}{rgb}{1.00,0.41,0.71}
\definecolor{IndianRed1}{rgb}{1.00,0.42,0.42}
\definecolor{IndianRed2}{rgb}{0.93,0.39,0.39}
\definecolor{IndianRed3}{rgb}{0.80,0.33,0.33}
\definecolor{IndianRed4}{rgb}{0.55,0.23,0.23}
\definecolor{IndianRed}{rgb}{0.80,0.36,0.36}
\definecolor{LavenderBlush1}{rgb}{1.00,0.94,0.96}
\definecolor{LavenderBlush2}{rgb}{0.93,0.88,0.90}
\definecolor{LavenderBlush3}{rgb}{0.80,0.76,0.77}
\definecolor{LavenderBlush4}{rgb}{0.55,0.51,0.53}
\definecolor{LavenderBlush}{rgb}{1.00,0.94,0.96}
\definecolor{LawnGreen}{rgb}{0.49,0.99,0.00}
\definecolor{LemonChiffon1}{rgb}{1.00,0.98,0.80}
\definecolor{LemonChiffon2}{rgb}{0.93,0.91,0.75}
\definecolor{LemonChiffon3}{rgb}{0.80,0.79,0.65}
\definecolor{LemonChiffon4}{rgb}{0.55,0.54,0.44}
\definecolor{LemonChiffon}{rgb}{1.00,0.98,0.80}
\definecolor{LightBlue1}{rgb}{0.75,0.94,1.00}
\definecolor{LightBlue2}{rgb}{0.70,0.87,0.93}
\definecolor{LightBlue3}{rgb}{0.60,0.75,0.80}
\definecolor{LightBlue4}{rgb}{0.41,0.51,0.55}
\definecolor{LightBlue}{rgb}{0.68,0.85,0.90}
\definecolor{LightCoral}{rgb}{0.94,0.50,0.50}
\definecolor{LightCyan1}{rgb}{0.88,1.00,1.00}
\definecolor{LightCyan2}{rgb}{0.82,0.93,0.93}
\definecolor{LightCyan3}{rgb}{0.71,0.80,0.80}
\definecolor{LightCyan4}{rgb}{0.48,0.55,0.55}
\definecolor{LightCyan}{rgb}{0.88,1.00,1.00}
\definecolor{LightGoldenrod1}{rgb}{1.00,0.93,0.55}
\definecolor{LightGoldenrod2}{rgb}{0.93,0.86,0.51}
\definecolor{LightGoldenrod3}{rgb}{0.80,0.75,0.44}
\definecolor{LightGoldenrod4}{rgb}{0.55,0.51,0.30}
\definecolor{LightGoldenrodYellow}{rgb}{0.98,0.98,0.82}
\definecolor{LightGoldenrod}{rgb}{0.93,0.87,0.51}
\definecolor{LightGray}{rgb}{0.83,0.83,0.83}
\definecolor{LightGreen}{rgb}{0.56,0.93,0.56}
\definecolor{LightGrey}{rgb}{0.83,0.83,0.83}
\definecolor{LightPink1}{rgb}{1.00,0.68,0.73}
\definecolor{LightPink2}{rgb}{0.93,0.64,0.68}
\definecolor{LightPink3}{rgb}{0.80,0.55,0.58}
\definecolor{LightPink4}{rgb}{0.55,0.37,0.40}
\definecolor{LightPink}{rgb}{1.00,0.71,0.76}
\definecolor{LightSalmon1}{rgb}{1.00,0.63,0.48}
\definecolor{LightSalmon2}{rgb}{0.93,0.58,0.45}
\definecolor{LightSalmon3}{rgb}{0.80,0.51,0.38}
\definecolor{LightSalmon4}{rgb}{0.55,0.34,0.26}
\definecolor{LightSalmon}{rgb}{1.00,0.63,0.48}
\definecolor{LightSeaGreen}{rgb}{0.13,0.70,0.67}
\definecolor{LightSkyBlue1}{rgb}{0.69,0.89,1.00}
\definecolor{LightSkyBlue2}{rgb}{0.64,0.83,0.93}
\definecolor{LightSkyBlue3}{rgb}{0.55,0.71,0.80}
\definecolor{LightSkyBlue4}{rgb}{0.38,0.48,0.55}
\definecolor{LightSkyBlue}{rgb}{0.53,0.81,0.98}
\definecolor{LightSlateBlue}{rgb}{0.52,0.44,1.00}
\definecolor{LightSlateGray}{rgb}{0.47,0.53,0.60}
\definecolor{LightSlateGrey}{rgb}{0.47,0.53,0.60}
\definecolor{LightSteelBlue1}{rgb}{0.79,0.88,1.00}
\definecolor{LightSteelBlue2}{rgb}{0.74,0.82,0.93}
\definecolor{LightSteelBlue3}{rgb}{0.64,0.71,0.80}
\definecolor{LightSteelBlue4}{rgb}{0.43,0.48,0.55}
\definecolor{LightSteelBlue}{rgb}{0.69,0.77,0.87}
\definecolor{LightYellow1}{rgb}{1.00,1.00,0.88}
\definecolor{LightYellow2}{rgb}{0.93,0.93,0.82}
\definecolor{LightYellow3}{rgb}{0.80,0.80,0.71}
\definecolor{LightYellow4}{rgb}{0.55,0.55,0.48}
\definecolor{LightYellow}{rgb}{1.00,1.00,0.88}
\definecolor{LimeGreen}{rgb}{0.20,0.80,0.20}
\definecolor{MediumAquamarine}{rgb}{0.40,0.80,0.67}
\definecolor{MediumBlue}{rgb}{0.00,0.00,0.80}
\definecolor{MediumOrchid1}{rgb}{0.88,0.40,1.00}
\definecolor{MediumOrchid2}{rgb}{0.82,0.37,0.93}
\definecolor{MediumOrchid3}{rgb}{0.71,0.32,0.80}
\definecolor{MediumOrchid4}{rgb}{0.48,0.22,0.55}
\definecolor{MediumOrchid}{rgb}{0.73,0.33,0.83}
\definecolor{MediumPurple1}{rgb}{0.67,0.51,1.00}
\definecolor{MediumPurple2}{rgb}{0.62,0.47,0.93}
\definecolor{MediumPurple3}{rgb}{0.54,0.41,0.80}
\definecolor{MediumPurple4}{rgb}{0.36,0.28,0.55}
\definecolor{MediumPurple}{rgb}{0.58,0.44,0.86}
\definecolor{MediumSeaGreen}{rgb}{0.24,0.70,0.44}
\definecolor{MediumSlateBlue}{rgb}{0.48,0.41,0.93}
\definecolor{MediumSpringGreen}{rgb}{0.00,0.98,0.60}
\definecolor{MediumTurquoise}{rgb}{0.28,0.82,0.80}
\definecolor{MediumVioletRed}{rgb}{0.78,0.08,0.52}
\definecolor{MidnightBlue}{rgb}{0.10,0.10,0.44}
\definecolor{MintCream}{rgb}{0.96,1.00,0.98}
\definecolor{MistyRose1}{rgb}{1.00,0.89,0.88}
\definecolor{MistyRose2}{rgb}{0.93,0.84,0.82}
\definecolor{MistyRose3}{rgb}{0.80,0.72,0.71}
\definecolor{MistyRose4}{rgb}{0.55,0.49,0.48}
\definecolor{MistyRose}{rgb}{1.00,0.89,0.88}
\definecolor{NavajoWhite1}{rgb}{1.00,0.87,0.68}
\definecolor{NavajoWhite2}{rgb}{0.93,0.81,0.63}
\definecolor{NavajoWhite3}{rgb}{0.80,0.70,0.55}
\definecolor{NavajoWhite4}{rgb}{0.55,0.47,0.37}
\definecolor{NavajoWhite}{rgb}{1.00,0.87,0.68}
\definecolor{NavyBlue}{rgb}{0.00,0.00,0.50}
\definecolor{OldLace}{rgb}{0.99,0.96,0.90}
\definecolor{OliveDrab1}{rgb}{0.75,1.00,0.24}
\definecolor{OliveDrab2}{rgb}{0.70,0.93,0.23}
\definecolor{OliveDrab3}{rgb}{0.60,0.80,0.20}
\definecolor{OliveDrab4}{rgb}{0.41,0.55,0.13}
\definecolor{OliveDrab}{rgb}{0.42,0.56,0.14}
\definecolor{OrangeRed1}{rgb}{1.00,0.27,0.00}
\definecolor{OrangeRed2}{rgb}{0.93,0.25,0.00}
\definecolor{OrangeRed3}{rgb}{0.80,0.22,0.00}
\definecolor{OrangeRed4}{rgb}{0.55,0.15,0.00}
\definecolor{OrangeRed}{rgb}{1.00,0.27,0.00}
\definecolor{PaleGoldenrod}{rgb}{0.93,0.91,0.67}
\definecolor{PaleGreen1}{rgb}{0.60,1.00,0.60}
\definecolor{PaleGreen2}{rgb}{0.56,0.93,0.56}
\definecolor{PaleGreen3}{rgb}{0.49,0.80,0.49}
\definecolor{PaleGreen4}{rgb}{0.33,0.55,0.33}
\definecolor{PaleGreen}{rgb}{0.60,0.98,0.60}
\definecolor{PaleTurquoise1}{rgb}{0.73,1.00,1.00}
\definecolor{PaleTurquoise2}{rgb}{0.68,0.93,0.93}
\definecolor{PaleTurquoise3}{rgb}{0.59,0.80,0.80}
\definecolor{PaleTurquoise4}{rgb}{0.40,0.55,0.55}
\definecolor{PaleTurquoise}{rgb}{0.69,0.93,0.93}
\definecolor{PaleVioletRed1}{rgb}{1.00,0.51,0.67}
\definecolor{PaleVioletRed2}{rgb}{0.93,0.47,0.62}
\definecolor{PaleVioletRed3}{rgb}{0.80,0.41,0.54}
\definecolor{PaleVioletRed4}{rgb}{0.55,0.28,0.36}
\definecolor{PaleVioletRed}{rgb}{0.86,0.44,0.58}
\definecolor{PapayaWhip}{rgb}{1.00,0.94,0.84}
\definecolor{PeachPuff1}{rgb}{1.00,0.85,0.73}
\definecolor{PeachPuff2}{rgb}{0.93,0.80,0.68}
\definecolor{PeachPuff3}{rgb}{0.80,0.69,0.58}
\definecolor{PeachPuff4}{rgb}{0.55,0.47,0.40}
\definecolor{PeachPuff}{rgb}{1.00,0.85,0.73}
\definecolor{PowderBlue}{rgb}{0.69,0.88,0.90}
\definecolor{RosyBrown1}{rgb}{1.00,0.76,0.76}
\definecolor{RosyBrown2}{rgb}{0.93,0.71,0.71}
\definecolor{RosyBrown3}{rgb}{0.80,0.61,0.61}
\definecolor{RosyBrown4}{rgb}{0.55,0.41,0.41}
\definecolor{RosyBrown}{rgb}{0.74,0.56,0.56}
\definecolor{RoyalBlue1}{rgb}{0.28,0.46,1.00}
\definecolor{RoyalBlue2}{rgb}{0.26,0.43,0.93}
\definecolor{RoyalBlue3}{rgb}{0.23,0.37,0.80}
\definecolor{RoyalBlue4}{rgb}{0.15,0.25,0.55}
\definecolor{RoyalBlue}{rgb}{0.25,0.41,0.88}
\definecolor{SaddleBrown}{rgb}{0.55,0.27,0.07}
\definecolor{SandyBrown}{rgb}{0.96,0.64,0.38}
\definecolor{SeaGreen1}{rgb}{0.33,1.00,0.62}
\definecolor{SeaGreen2}{rgb}{0.31,0.93,0.58}
\definecolor{SeaGreen3}{rgb}{0.26,0.80,0.50}
\definecolor{SeaGreen4}{rgb}{0.18,0.55,0.34}
\definecolor{SeaGreen}{rgb}{0.18,0.55,0.34}
\definecolor{SkyBlue1}{rgb}{0.53,0.81,1.00}
\definecolor{SkyBlue2}{rgb}{0.49,0.75,0.93}
\definecolor{SkyBlue3}{rgb}{0.42,0.65,0.80}
\definecolor{SkyBlue4}{rgb}{0.29,0.44,0.55}
\definecolor{SkyBlue}{rgb}{0.53,0.81,0.92}
\definecolor{SlateBlue1}{rgb}{0.51,0.44,1.00}
\definecolor{SlateBlue2}{rgb}{0.48,0.40,0.93}
\definecolor{SlateBlue3}{rgb}{0.41,0.35,0.80}
\definecolor{SlateBlue4}{rgb}{0.28,0.24,0.55}
\definecolor{SlateBlue}{rgb}{0.42,0.35,0.80}
\definecolor{SlateGray1}{rgb}{0.78,0.89,1.00}
\definecolor{SlateGray2}{rgb}{0.73,0.83,0.93}
\definecolor{SlateGray3}{rgb}{0.62,0.71,0.80}
\definecolor{SlateGray4}{rgb}{0.42,0.48,0.55}
\definecolor{SlateGray}{rgb}{0.44,0.50,0.56}
\definecolor{SlateGrey}{rgb}{0.44,0.50,0.56}
\definecolor{SpringGreen1}{rgb}{0.00,1.00,0.50}
\definecolor{SpringGreen2}{rgb}{0.00,0.93,0.46}
\definecolor{SpringGreen3}{rgb}{0.00,0.80,0.40}
\definecolor{SpringGreen4}{rgb}{0.00,0.55,0.27}
\definecolor{SpringGreen}{rgb}{0.00,1.00,0.50}
\definecolor{SteelBlue1}{rgb}{0.39,0.72,1.00}
\definecolor{SteelBlue2}{rgb}{0.36,0.67,0.93}
\definecolor{SteelBlue3}{rgb}{0.31,0.58,0.80}
\definecolor{SteelBlue4}{rgb}{0.21,0.39,0.55}
\definecolor{SteelBlue}{rgb}{0.27,0.51,0.71}
\definecolor{VioletRed1}{rgb}{1.00,0.24,0.59}
\definecolor{VioletRed2}{rgb}{0.93,0.23,0.55}
\definecolor{VioletRed3}{rgb}{0.80,0.20,0.47}
\definecolor{VioletRed4}{rgb}{0.55,0.13,0.32}
\definecolor{VioletRed}{rgb}{0.82,0.13,0.56}
\definecolor{WhiteSmoke}{rgb}{0.96,0.96,0.96}
\definecolor{YellowGreen}{rgb}{0.60,0.80,0.20}
\definecolor{aliceblue}{rgb}{0.94,0.97,1.00}
\definecolor{antiquewhite}{rgb}{0.98,0.92,0.84}
\definecolor{aquamarine1}{rgb}{0.50,1.00,0.83}
\definecolor{aquamarine2}{rgb}{0.46,0.93,0.78}
\definecolor{aquamarine3}{rgb}{0.40,0.80,0.67}
\definecolor{aquamarine4}{rgb}{0.27,0.55,0.45}
\definecolor{aquamarine}{rgb}{0.50,1.00,0.83}
\definecolor{azure1}{rgb}{0.94,1.00,1.00}
\definecolor{azure2}{rgb}{0.88,0.93,0.93}
\definecolor{azure3}{rgb}{0.76,0.80,0.80}
\definecolor{azure4}{rgb}{0.51,0.55,0.55}
\definecolor{azure}{rgb}{0.94,1.00,1.00}
\definecolor{beige}{rgb}{0.96,0.96,0.86}
\definecolor{bisque1}{rgb}{1.00,0.89,0.77}
\definecolor{bisque2}{rgb}{0.93,0.84,0.72}
\definecolor{bisque3}{rgb}{0.80,0.72,0.62}
\definecolor{bisque4}{rgb}{0.55,0.49,0.42}
\definecolor{bisque}{rgb}{1.00,0.89,0.77}
\definecolor{black}{rgb}{0.00,0.00,0.00}
\definecolor{blanchedalmond}{rgb}{1.00,0.92,0.80}
\definecolor{blue1}{rgb}{0.00,0.00,1.00}
\definecolor{blue2}{rgb}{0.00,0.00,0.93}
\definecolor{blue3}{rgb}{0.00,0.00,0.80}
\definecolor{blue4}{rgb}{0.00,0.00,0.55}
\definecolor{blueviolet}{rgb}{0.54,0.17,0.89}
\definecolor{blue}{rgb}{0.00,0.00,1.00}
\definecolor{brown1}{rgb}{1.00,0.25,0.25}
\definecolor{brown2}{rgb}{0.93,0.23,0.23}
\definecolor{brown3}{rgb}{0.80,0.20,0.20}
\definecolor{brown4}{rgb}{0.55,0.14,0.14}
\definecolor{brown}{rgb}{0.65,0.16,0.16}
\definecolor{burlywood1}{rgb}{1.00,0.83,0.61}
\definecolor{burlywood2}{rgb}{0.93,0.77,0.57}
\definecolor{burlywood3}{rgb}{0.80,0.67,0.49}
\definecolor{burlywood4}{rgb}{0.55,0.45,0.33}
\definecolor{burlywood}{rgb}{0.87,0.72,0.53}
\definecolor{cadetblue}{rgb}{0.37,0.62,0.63}
\definecolor{chartreuse1}{rgb}{0.50,1.00,0.00}
\definecolor{chartreuse2}{rgb}{0.46,0.93,0.00}
\definecolor{chartreuse3}{rgb}{0.40,0.80,0.00}
\definecolor{chartreuse4}{rgb}{0.27,0.55,0.00}
\definecolor{chartreuse}{rgb}{0.50,1.00,0.00}
\definecolor{chocolate1}{rgb}{1.00,0.50,0.14}
\definecolor{chocolate2}{rgb}{0.93,0.46,0.13}
\definecolor{chocolate3}{rgb}{0.80,0.40,0.11}
\definecolor{chocolate4}{rgb}{0.55,0.27,0.07}
\definecolor{chocolate}{rgb}{0.82,0.41,0.12}
\definecolor{coral1}{rgb}{1.00,0.45,0.34}
\definecolor{coral2}{rgb}{0.93,0.42,0.31}
\definecolor{coral3}{rgb}{0.80,0.36,0.27}
\definecolor{coral4}{rgb}{0.55,0.24,0.18}
\definecolor{coral}{rgb}{1.00,0.50,0.31}
\definecolor{cornflowerblue}{rgb}{0.39,0.58,0.93}
\definecolor{cornsilk1}{rgb}{1.00,0.97,0.86}
\definecolor{cornsilk2}{rgb}{0.93,0.91,0.80}
\definecolor{cornsilk3}{rgb}{0.80,0.78,0.69}
\definecolor{cornsilk4}{rgb}{0.55,0.53,0.47}
\definecolor{cornsilk}{rgb}{1.00,0.97,0.86}
\definecolor{cyan1}{rgb}{0.00,1.00,1.00}
\definecolor{cyan2}{rgb}{0.00,0.93,0.93}
\definecolor{cyan3}{rgb}{0.00,0.80,0.80}
\definecolor{cyan4}{rgb}{0.00,0.55,0.55}
\definecolor{cyan}{rgb}{0.00,1.00,1.00}
\definecolor{darkblue}{rgb}{0.00,0.00,0.55}
\definecolor{darkcyan}{rgb}{0.00,0.55,0.55}
\definecolor{darkgoldenrod}{rgb}{0.72,0.53,0.04}
\definecolor{darkgray}{rgb}{0.66,0.66,0.66}
\definecolor{darkgreen}{rgb}{0.00,0.39,0.00}
\definecolor{darkgrey}{rgb}{0.66,0.66,0.66}
\definecolor{darkkhaki}{rgb}{0.74,0.72,0.42}
\definecolor{darkmagenta}{rgb}{0.55,0.00,0.55}
\definecolor{darkolive}{rgb}{0.33,0.42,0.18}
\definecolor{darkorange}{rgb}{1.00,0.55,0.00}
\definecolor{darkorchid}{rgb}{0.60,0.20,0.80}
\definecolor{darkred}{rgb}{0.55,0.00,0.00}
\definecolor{darksalmon}{rgb}{0.91,0.59,0.48}
\definecolor{darksea}{rgb}{0.56,0.74,0.56}
\definecolor{darkslate}{rgb}{0.18,0.31,0.31}
\definecolor{darkslate}{rgb}{0.18,0.31,0.31}
\definecolor{darkslate}{rgb}{0.28,0.24,0.55}
\definecolor{darkturquoise}{rgb}{0.00,0.81,0.82}
\definecolor{darkviolet}{rgb}{0.58,0.00,0.83}
\definecolor{deeppink}{rgb}{1.00,0.08,0.58}
\definecolor{deepsky}{rgb}{0.00,0.75,1.00}
\definecolor{dimgray}{rgb}{0.41,0.41,0.41}
\definecolor{dimgrey}{rgb}{0.41,0.41,0.41}
\definecolor{dodgerblue}{rgb}{0.12,0.56,1.00}
\definecolor{firebrick1}{rgb}{1.00,0.19,0.19}
\definecolor{firebrick2}{rgb}{0.93,0.17,0.17}
\definecolor{firebrick3}{rgb}{0.80,0.15,0.15}
\definecolor{firebrick4}{rgb}{0.55,0.10,0.10}
\definecolor{firebrick}{rgb}{0.70,0.13,0.13}
\definecolor{floralwhite}{rgb}{1.00,0.98,0.94}
\definecolor{forestgreen}{rgb}{0.13,0.55,0.13}
\definecolor{gainsboro}{rgb}{0.86,0.86,0.86}
\definecolor{ghostwhite}{rgb}{0.97,0.97,1.00}
\definecolor{gold1}{rgb}{1.00,0.84,0.00}
\definecolor{gold2}{rgb}{0.93,0.79,0.00}
\definecolor{gold3}{rgb}{0.80,0.68,0.00}
\definecolor{gold4}{rgb}{0.55,0.46,0.00}
\definecolor{goldenrod1}{rgb}{1.00,0.76,0.15}
\definecolor{goldenrod2}{rgb}{0.93,0.71,0.13}
\definecolor{goldenrod3}{rgb}{0.80,0.61,0.11}
\definecolor{goldenrod4}{rgb}{0.55,0.41,0.08}
\definecolor{goldenrod}{rgb}{0.85,0.65,0.13}
\definecolor{gold}{rgb}{1.00,0.84,0.00}
\definecolor{gray0}{rgb}{0.00,0.00,0.00}
\definecolor{gray100}{rgb}{1.00,1.00,1.00}
\definecolor{gray10}{rgb}{0.10,0.10,0.10}
\definecolor{gray11}{rgb}{0.11,0.11,0.11}
\definecolor{gray12}{rgb}{0.12,0.12,0.12}
\definecolor{gray13}{rgb}{0.13,0.13,0.13}
\definecolor{gray14}{rgb}{0.14,0.14,0.14}
\definecolor{gray15}{rgb}{0.15,0.15,0.15}
\definecolor{gray16}{rgb}{0.16,0.16,0.16}
\definecolor{gray17}{rgb}{0.17,0.17,0.17}
\definecolor{gray18}{rgb}{0.18,0.18,0.18}
\definecolor{gray19}{rgb}{0.19,0.19,0.19}
\definecolor{gray1}{rgb}{0.01,0.01,0.01}
\definecolor{gray20}{rgb}{0.20,0.20,0.20}
\definecolor{gray21}{rgb}{0.21,0.21,0.21}
\definecolor{gray22}{rgb}{0.22,0.22,0.22}
\definecolor{gray23}{rgb}{0.23,0.23,0.23}
\definecolor{gray24}{rgb}{0.24,0.24,0.24}
\definecolor{gray25}{rgb}{0.25,0.25,0.25}
\definecolor{gray26}{rgb}{0.26,0.26,0.26}
\definecolor{gray27}{rgb}{0.27,0.27,0.27}
\definecolor{gray28}{rgb}{0.28,0.28,0.28}
\definecolor{gray29}{rgb}{0.29,0.29,0.29}
\definecolor{gray2}{rgb}{0.02,0.02,0.02}
\definecolor{gray30}{rgb}{0.30,0.30,0.30}
\definecolor{gray31}{rgb}{0.31,0.31,0.31}
\definecolor{gray32}{rgb}{0.32,0.32,0.32}
\definecolor{gray33}{rgb}{0.33,0.33,0.33}
\definecolor{gray34}{rgb}{0.34,0.34,0.34}
\definecolor{gray35}{rgb}{0.35,0.35,0.35}
\definecolor{gray36}{rgb}{0.36,0.36,0.36}
\definecolor{gray37}{rgb}{0.37,0.37,0.37}
\definecolor{gray38}{rgb}{0.38,0.38,0.38}
\definecolor{gray39}{rgb}{0.39,0.39,0.39}
\definecolor{gray3}{rgb}{0.03,0.03,0.03}
\definecolor{gray40}{rgb}{0.40,0.40,0.40}
\definecolor{gray41}{rgb}{0.41,0.41,0.41}
\definecolor{gray42}{rgb}{0.42,0.42,0.42}
\definecolor{gray43}{rgb}{0.43,0.43,0.43}
\definecolor{gray44}{rgb}{0.44,0.44,0.44}
\definecolor{gray45}{rgb}{0.45,0.45,0.45}
\definecolor{gray46}{rgb}{0.46,0.46,0.46}
\definecolor{gray47}{rgb}{0.47,0.47,0.47}
\definecolor{gray48}{rgb}{0.48,0.48,0.48}
\definecolor{gray49}{rgb}{0.49,0.49,0.49}
\definecolor{gray4}{rgb}{0.04,0.04,0.04}
\definecolor{gray50}{rgb}{0.50,0.50,0.50}
\definecolor{gray51}{rgb}{0.51,0.51,0.51}
\definecolor{gray52}{rgb}{0.52,0.52,0.52}
\definecolor{gray53}{rgb}{0.53,0.53,0.53}
\definecolor{gray54}{rgb}{0.54,0.54,0.54}
\definecolor{gray55}{rgb}{0.55,0.55,0.55}
\definecolor{gray56}{rgb}{0.56,0.56,0.56}
\definecolor{gray57}{rgb}{0.57,0.57,0.57}
\definecolor{gray58}{rgb}{0.58,0.58,0.58}
\definecolor{gray59}{rgb}{0.59,0.59,0.59}
\definecolor{gray5}{rgb}{0.05,0.05,0.05}
\definecolor{gray60}{rgb}{0.60,0.60,0.60}
\definecolor{gray61}{rgb}{0.61,0.61,0.61}
\definecolor{gray62}{rgb}{0.62,0.62,0.62}
\definecolor{gray63}{rgb}{0.63,0.63,0.63}
\definecolor{gray64}{rgb}{0.64,0.64,0.64}
\definecolor{gray65}{rgb}{0.65,0.65,0.65}
\definecolor{gray66}{rgb}{0.66,0.66,0.66}
\definecolor{gray67}{rgb}{0.67,0.67,0.67}
\definecolor{gray68}{rgb}{0.68,0.68,0.68}
\definecolor{gray69}{rgb}{0.69,0.69,0.69}
\definecolor{gray6}{rgb}{0.06,0.06,0.06}
\definecolor{gray70}{rgb}{0.70,0.70,0.70}
\definecolor{gray71}{rgb}{0.71,0.71,0.71}
\definecolor{gray72}{rgb}{0.72,0.72,0.72}
\definecolor{gray73}{rgb}{0.73,0.73,0.73}
\definecolor{gray74}{rgb}{0.74,0.74,0.74}
\definecolor{gray75}{rgb}{0.75,0.75,0.75}
\definecolor{gray76}{rgb}{0.76,0.76,0.76}
\definecolor{gray77}{rgb}{0.77,0.77,0.77}
\definecolor{gray78}{rgb}{0.78,0.78,0.78}
\definecolor{gray79}{rgb}{0.79,0.79,0.79}
\definecolor{gray7}{rgb}{0.07,0.07,0.07}
\definecolor{gray80}{rgb}{0.80,0.80,0.80}
\definecolor{gray81}{rgb}{0.81,0.81,0.81}
\definecolor{gray82}{rgb}{0.82,0.82,0.82}
\definecolor{gray83}{rgb}{0.83,0.83,0.83}
\definecolor{gray84}{rgb}{0.84,0.84,0.84}
\definecolor{gray85}{rgb}{0.85,0.85,0.85}
\definecolor{gray86}{rgb}{0.86,0.86,0.86}
\definecolor{gray87}{rgb}{0.87,0.87,0.87}
\definecolor{gray88}{rgb}{0.88,0.88,0.88}
\definecolor{gray89}{rgb}{0.89,0.89,0.89}
\definecolor{gray8}{rgb}{0.08,0.08,0.08}
\definecolor{gray90}{rgb}{0.90,0.90,0.90}
\definecolor{gray91}{rgb}{0.91,0.91,0.91}
\definecolor{gray92}{rgb}{0.92,0.92,0.92}
\definecolor{gray93}{rgb}{0.93,0.93,0.93}
\definecolor{gray94}{rgb}{0.94,0.94,0.94}
\definecolor{gray95}{rgb}{0.95,0.95,0.95}
\definecolor{gray96}{rgb}{0.96,0.96,0.96}
\definecolor{gray97}{rgb}{0.97,0.97,0.97}
\definecolor{gray98}{rgb}{0.98,0.98,0.98}
\definecolor{gray99}{rgb}{0.99,0.99,0.99}
\definecolor{gray9}{rgb}{0.09,0.09,0.09}
\definecolor{gray}{rgb}{0.75,0.75,0.75}
\definecolor{green1}{rgb}{0.00,1.00,0.00}
\definecolor{green2}{rgb}{0.00,0.93,0.00}
\definecolor{green3}{rgb}{0.00,0.80,0.00}
\definecolor{green4}{rgb}{0.00,0.55,0.00}
\definecolor{greenyellow}{rgb}{0.68,1.00,0.18}
\definecolor{green}{rgb}{0.00,1.00,0.00}
\definecolor{grey0}{rgb}{0.00,0.00,0.00}
\definecolor{grey100}{rgb}{1.00,1.00,1.00}
\definecolor{grey10}{rgb}{0.10,0.10,0.10}
\definecolor{grey11}{rgb}{0.11,0.11,0.11}
\definecolor{grey12}{rgb}{0.12,0.12,0.12}
\definecolor{grey13}{rgb}{0.13,0.13,0.13}
\definecolor{grey14}{rgb}{0.14,0.14,0.14}
\definecolor{grey15}{rgb}{0.15,0.15,0.15}
\definecolor{grey16}{rgb}{0.16,0.16,0.16}
\definecolor{grey17}{rgb}{0.17,0.17,0.17}
\definecolor{grey18}{rgb}{0.18,0.18,0.18}
\definecolor{grey19}{rgb}{0.19,0.19,0.19}
\definecolor{grey1}{rgb}{0.01,0.01,0.01}
\definecolor{grey20}{rgb}{0.20,0.20,0.20}
\definecolor{grey21}{rgb}{0.21,0.21,0.21}
\definecolor{grey22}{rgb}{0.22,0.22,0.22}
\definecolor{grey23}{rgb}{0.23,0.23,0.23}
\definecolor{grey24}{rgb}{0.24,0.24,0.24}
\definecolor{grey25}{rgb}{0.25,0.25,0.25}
\definecolor{grey26}{rgb}{0.26,0.26,0.26}
\definecolor{grey27}{rgb}{0.27,0.27,0.27}
\definecolor{grey28}{rgb}{0.28,0.28,0.28}
\definecolor{grey29}{rgb}{0.29,0.29,0.29}
\definecolor{grey2}{rgb}{0.02,0.02,0.02}
\definecolor{grey30}{rgb}{0.30,0.30,0.30}
\definecolor{grey31}{rgb}{0.31,0.31,0.31}
\definecolor{grey32}{rgb}{0.32,0.32,0.32}
\definecolor{grey33}{rgb}{0.33,0.33,0.33}
\definecolor{grey34}{rgb}{0.34,0.34,0.34}
\definecolor{grey35}{rgb}{0.35,0.35,0.35}
\definecolor{grey36}{rgb}{0.36,0.36,0.36}
\definecolor{grey37}{rgb}{0.37,0.37,0.37}
\definecolor{grey38}{rgb}{0.38,0.38,0.38}
\definecolor{grey39}{rgb}{0.39,0.39,0.39}
\definecolor{grey3}{rgb}{0.03,0.03,0.03}
\definecolor{grey40}{rgb}{0.40,0.40,0.40}
\definecolor{grey41}{rgb}{0.41,0.41,0.41}
\definecolor{grey42}{rgb}{0.42,0.42,0.42}
\definecolor{grey43}{rgb}{0.43,0.43,0.43}
\definecolor{grey44}{rgb}{0.44,0.44,0.44}
\definecolor{grey45}{rgb}{0.45,0.45,0.45}
\definecolor{grey46}{rgb}{0.46,0.46,0.46}
\definecolor{grey47}{rgb}{0.47,0.47,0.47}
\definecolor{grey48}{rgb}{0.48,0.48,0.48}
\definecolor{grey49}{rgb}{0.49,0.49,0.49}
\definecolor{grey4}{rgb}{0.04,0.04,0.04}
\definecolor{grey50}{rgb}{0.50,0.50,0.50}
\definecolor{grey51}{rgb}{0.51,0.51,0.51}
\definecolor{grey52}{rgb}{0.52,0.52,0.52}
\definecolor{grey53}{rgb}{0.53,0.53,0.53}
\definecolor{grey54}{rgb}{0.54,0.54,0.54}
\definecolor{grey55}{rgb}{0.55,0.55,0.55}
\definecolor{grey56}{rgb}{0.56,0.56,0.56}
\definecolor{grey57}{rgb}{0.57,0.57,0.57}
\definecolor{grey58}{rgb}{0.58,0.58,0.58}
\definecolor{grey59}{rgb}{0.59,0.59,0.59}
\definecolor{grey5}{rgb}{0.05,0.05,0.05}
\definecolor{grey60}{rgb}{0.60,0.60,0.60}
\definecolor{grey61}{rgb}{0.61,0.61,0.61}
\definecolor{grey62}{rgb}{0.62,0.62,0.62}
\definecolor{grey63}{rgb}{0.63,0.63,0.63}
\definecolor{grey64}{rgb}{0.64,0.64,0.64}
\definecolor{grey65}{rgb}{0.65,0.65,0.65}
\definecolor{grey66}{rgb}{0.66,0.66,0.66}
\definecolor{grey67}{rgb}{0.67,0.67,0.67}
\definecolor{grey68}{rgb}{0.68,0.68,0.68}
\definecolor{grey69}{rgb}{0.69,0.69,0.69}
\definecolor{grey6}{rgb}{0.06,0.06,0.06}
\definecolor{grey70}{rgb}{0.70,0.70,0.70}
\definecolor{grey71}{rgb}{0.71,0.71,0.71}
\definecolor{grey72}{rgb}{0.72,0.72,0.72}
\definecolor{grey73}{rgb}{0.73,0.73,0.73}
\definecolor{grey74}{rgb}{0.74,0.74,0.74}
\definecolor{grey75}{rgb}{0.75,0.75,0.75}
\definecolor{grey76}{rgb}{0.76,0.76,0.76}
\definecolor{grey77}{rgb}{0.77,0.77,0.77}
\definecolor{grey78}{rgb}{0.78,0.78,0.78}
\definecolor{grey79}{rgb}{0.79,0.79,0.79}
\definecolor{grey7}{rgb}{0.07,0.07,0.07}
\definecolor{grey80}{rgb}{0.80,0.80,0.80}
\definecolor{grey81}{rgb}{0.81,0.81,0.81}
\definecolor{grey82}{rgb}{0.82,0.82,0.82}
\definecolor{grey83}{rgb}{0.83,0.83,0.83}
\definecolor{grey84}{rgb}{0.84,0.84,0.84}
\definecolor{grey85}{rgb}{0.85,0.85,0.85}
\definecolor{grey86}{rgb}{0.86,0.86,0.86}
\definecolor{grey87}{rgb}{0.87,0.87,0.87}
\definecolor{grey88}{rgb}{0.88,0.88,0.88}
\definecolor{grey89}{rgb}{0.89,0.89,0.89}
\definecolor{grey8}{rgb}{0.08,0.08,0.08}
\definecolor{grey90}{rgb}{0.90,0.90,0.90}
\definecolor{grey91}{rgb}{0.91,0.91,0.91}
\definecolor{grey92}{rgb}{0.92,0.92,0.92}
\definecolor{grey93}{rgb}{0.93,0.93,0.93}
\definecolor{grey94}{rgb}{0.94,0.94,0.94}
\definecolor{grey95}{rgb}{0.95,0.95,0.95}
\definecolor{grey96}{rgb}{0.96,0.96,0.96}
\definecolor{grey97}{rgb}{0.97,0.97,0.97}
\definecolor{grey98}{rgb}{0.98,0.98,0.98}
\definecolor{grey99}{rgb}{0.99,0.99,0.99}
\definecolor{grey9}{rgb}{0.09,0.09,0.09}
\definecolor{grey}{rgb}{0.75,0.75,0.75}
\definecolor{honeydew1}{rgb}{0.94,1.00,0.94}
\definecolor{honeydew2}{rgb}{0.88,0.93,0.88}
\definecolor{honeydew3}{rgb}{0.76,0.80,0.76}
\definecolor{honeydew4}{rgb}{0.51,0.55,0.51}
\definecolor{honeydew}{rgb}{0.94,1.00,0.94}
\definecolor{hotpink}{rgb}{1.00,0.41,0.71}
\definecolor{indianred}{rgb}{0.80,0.36,0.36}
\definecolor{ivory1}{rgb}{1.00,1.00,0.94}
\definecolor{ivory2}{rgb}{0.93,0.93,0.88}
\definecolor{ivory3}{rgb}{0.80,0.80,0.76}
\definecolor{ivory4}{rgb}{0.55,0.55,0.51}
\definecolor{ivory}{rgb}{1.00,1.00,0.94}
\definecolor{khaki1}{rgb}{1.00,0.96,0.56}
\definecolor{khaki2}{rgb}{0.93,0.90,0.52}
\definecolor{khaki3}{rgb}{0.80,0.78,0.45}
\definecolor{khaki4}{rgb}{0.55,0.53,0.31}
\definecolor{khaki}{rgb}{0.94,0.90,0.55}
\definecolor{lavenderblush}{rgb}{1.00,0.94,0.96}
\definecolor{lavender}{rgb}{0.90,0.90,0.98}
\definecolor{lawngreen}{rgb}{0.49,0.99,0.00}
\definecolor{lemonchiffon}{rgb}{1.00,0.98,0.80}
\definecolor{lightblue}{rgb}{0.68,0.85,0.90}
\definecolor{lightcoral}{rgb}{0.94,0.50,0.50}
\definecolor{lightcyan}{rgb}{0.88,1.00,1.00}
\definecolor{lightgoldenrod}{rgb}{0.93,0.87,0.51}
\definecolor{lightgoldenrod}{rgb}{0.98,0.98,0.82}
\definecolor{lightgray}{rgb}{0.83,0.83,0.83}
\definecolor{lightgreen}{rgb}{0.56,0.93,0.56}
\definecolor{lightgrey}{rgb}{0.83,0.83,0.83}
\definecolor{lightpink}{rgb}{1.00,0.71,0.76}
\definecolor{lightsalmon}{rgb}{1.00,0.63,0.48}
\definecolor{lightsea}{rgb}{0.13,0.70,0.67}
\definecolor{lightsky}{rgb}{0.53,0.81,0.98}
\definecolor{lightslate}{rgb}{0.47,0.53,0.60}
\definecolor{lightslate}{rgb}{0.47,0.53,0.60}
\definecolor{lightslate}{rgb}{0.52,0.44,1.00}
\definecolor{lightsteel}{rgb}{0.69,0.77,0.87}
\definecolor{lightyellow}{rgb}{1.00,1.00,0.88}
\definecolor{limegreen}{rgb}{0.20,0.80,0.20}
\definecolor{linen}{rgb}{0.98,0.94,0.90}
\definecolor{magenta1}{rgb}{1.00,0.00,1.00}
\definecolor{magenta2}{rgb}{0.93,0.00,0.93}
\definecolor{magenta3}{rgb}{0.80,0.00,0.80}
\definecolor{magenta4}{rgb}{0.55,0.00,0.55}
\definecolor{magenta}{rgb}{1.00,0.00,1.00}
\definecolor{maroon1}{rgb}{1.00,0.20,0.70}
\definecolor{maroon2}{rgb}{0.93,0.19,0.65}
\definecolor{maroon3}{rgb}{0.80,0.16,0.56}
\definecolor{maroon4}{rgb}{0.55,0.11,0.38}
\definecolor{maroon}{rgb}{0.69,0.19,0.38}
\definecolor{mediumaquamarine}{rgb}{0.40,0.80,0.67}
\definecolor{mediumblue}{rgb}{0.00,0.00,0.80}
\definecolor{mediumorchid}{rgb}{0.73,0.33,0.83}
\definecolor{mediumpurple}{rgb}{0.58,0.44,0.86}
\definecolor{mediumsea}{rgb}{0.24,0.70,0.44}
\definecolor{mediumslate}{rgb}{0.48,0.41,0.93}
\definecolor{mediumspring}{rgb}{0.00,0.98,0.60}
\definecolor{mediumturquoise}{rgb}{0.28,0.82,0.80}
\definecolor{mediumviolet}{rgb}{0.78,0.08,0.52}
\definecolor{midnightblue}{rgb}{0.10,0.10,0.44}
\definecolor{mintcream}{rgb}{0.96,1.00,0.98}
\definecolor{mistyrose}{rgb}{1.00,0.89,0.88}
\definecolor{moccasin}{rgb}{1.00,0.89,0.71}
\definecolor{navajowhite}{rgb}{1.00,0.87,0.68}
\definecolor{navyblue}{rgb}{0.00,0.00,0.50}
\definecolor{navy}{rgb}{0.00,0.00,0.50}
\definecolor{oldlace}{rgb}{0.99,0.96,0.90}
\definecolor{olivedrab}{rgb}{0.42,0.56,0.14}
\definecolor{orange1}{rgb}{1.00,0.65,0.00}
\definecolor{orange2}{rgb}{0.93,0.60,0.00}
\definecolor{orange3}{rgb}{0.80,0.52,0.00}
\definecolor{orange4}{rgb}{0.55,0.35,0.00}
\definecolor{orangered}{rgb}{1.00,0.27,0.00}
\definecolor{orange}{rgb}{1.00,0.65,0.00}
\definecolor{orchid1}{rgb}{1.00,0.51,0.98}
\definecolor{orchid2}{rgb}{0.93,0.48,0.91}
\definecolor{orchid3}{rgb}{0.80,0.41,0.79}
\definecolor{orchid4}{rgb}{0.55,0.28,0.54}
\definecolor{orchid}{rgb}{0.85,0.44,0.84}
\definecolor{palegoldenrod}{rgb}{0.93,0.91,0.67}
\definecolor{palegreen}{rgb}{0.60,0.98,0.60}
\definecolor{paleturquoise}{rgb}{0.69,0.93,0.93}
\definecolor{paleviolet}{rgb}{0.86,0.44,0.58}
\definecolor{papayawhip}{rgb}{1.00,0.94,0.84}
\definecolor{peachpuff}{rgb}{1.00,0.85,0.73}
\definecolor{peru}{rgb}{0.80,0.52,0.25}
\definecolor{pink1}{rgb}{1.00,0.71,0.77}
\definecolor{pink2}{rgb}{0.93,0.66,0.72}
\definecolor{pink3}{rgb}{0.80,0.57,0.62}
\definecolor{pink4}{rgb}{0.55,0.39,0.42}
\definecolor{pink}{rgb}{1.00,0.75,0.80}
\definecolor{plum1}{rgb}{1.00,0.73,1.00}
\definecolor{plum2}{rgb}{0.93,0.68,0.93}
\definecolor{plum3}{rgb}{0.80,0.59,0.80}
\definecolor{plum4}{rgb}{0.55,0.40,0.55}
\definecolor{plum}{rgb}{0.87,0.63,0.87}
\definecolor{powderblue}{rgb}{0.69,0.88,0.90}
\definecolor{purple1}{rgb}{0.61,0.19,1.00}
\definecolor{purple2}{rgb}{0.57,0.17,0.93}
\definecolor{purple3}{rgb}{0.49,0.15,0.80}
\definecolor{purple4}{rgb}{0.33,0.10,0.55}
\definecolor{purple}{rgb}{0.63,0.13,0.94}
\definecolor{red1}{rgb}{1.00,0.00,0.00}
\definecolor{red2}{rgb}{0.93,0.00,0.00}
\definecolor{red3}{rgb}{0.80,0.00,0.00}
\definecolor{red4}{rgb}{0.55,0.00,0.00}
\definecolor{red}{rgb}{1.00,0.00,0.00}
\definecolor{rosybrown}{rgb}{0.74,0.56,0.56}
\definecolor{royalblue}{rgb}{0.25,0.41,0.88}
\definecolor{saddlebrown}{rgb}{0.55,0.27,0.07}
\definecolor{salmon1}{rgb}{1.00,0.55,0.41}
\definecolor{salmon2}{rgb}{0.93,0.51,0.38}
\definecolor{salmon3}{rgb}{0.80,0.44,0.33}
\definecolor{salmon4}{rgb}{0.55,0.30,0.22}
\definecolor{salmon}{rgb}{0.98,0.50,0.45}
\definecolor{sandybrown}{rgb}{0.96,0.64,0.38}
\definecolor{seagreen}{rgb}{0.18,0.55,0.34}
\definecolor{seashell1}{rgb}{1.00,0.96,0.93}
\definecolor{seashell2}{rgb}{0.93,0.90,0.87}
\definecolor{seashell3}{rgb}{0.80,0.77,0.75}
\definecolor{seashell4}{rgb}{0.55,0.53,0.51}
\definecolor{seashell}{rgb}{1.00,0.96,0.93}
\definecolor{sienna1}{rgb}{1.00,0.51,0.28}
\definecolor{sienna2}{rgb}{0.93,0.47,0.26}
\definecolor{sienna3}{rgb}{0.80,0.41,0.22}
\definecolor{sienna4}{rgb}{0.55,0.28,0.15}
\definecolor{sienna}{rgb}{0.63,0.32,0.18}
\definecolor{skyblue}{rgb}{0.53,0.81,0.92}
\definecolor{slateblue}{rgb}{0.42,0.35,0.80}
\definecolor{slategray}{rgb}{0.44,0.50,0.56}
\definecolor{slategrey}{rgb}{0.44,0.50,0.56}
\definecolor{snow1}{rgb}{1.00,0.98,0.98}
\definecolor{snow2}{rgb}{0.93,0.91,0.91}
\definecolor{snow3}{rgb}{0.80,0.79,0.79}
\definecolor{snow4}{rgb}{0.55,0.54,0.54}
\definecolor{snow}{rgb}{1.00,0.98,0.98}
\definecolor{springgreen}{rgb}{0.00,1.00,0.50}
\definecolor{steelblue}{rgb}{0.27,0.51,0.71}
\definecolor{tan1}{rgb}{1.00,0.65,0.31}
\definecolor{tan2}{rgb}{0.93,0.60,0.29}
\definecolor{tan3}{rgb}{0.80,0.52,0.25}
\definecolor{tan4}{rgb}{0.55,0.35,0.17}
\definecolor{tan}{rgb}{0.82,0.71,0.55}
\definecolor{thistle1}{rgb}{1.00,0.88,1.00}
\definecolor{thistle2}{rgb}{0.93,0.82,0.93}
\definecolor{thistle3}{rgb}{0.80,0.71,0.80}
\definecolor{thistle4}{rgb}{0.55,0.48,0.55}
\definecolor{thistle}{rgb}{0.85,0.75,0.85}
\definecolor{tomato1}{rgb}{1.00,0.39,0.28}
\definecolor{tomato2}{rgb}{0.93,0.36,0.26}
\definecolor{tomato3}{rgb}{0.80,0.31,0.22}
\definecolor{tomato4}{rgb}{0.55,0.21,0.15}
\definecolor{tomato}{rgb}{1.00,0.39,0.28}
\definecolor{turquoise1}{rgb}{0.00,0.96,1.00}
\definecolor{turquoise2}{rgb}{0.00,0.90,0.93}
\definecolor{turquoise3}{rgb}{0.00,0.77,0.80}
\definecolor{turquoise4}{rgb}{0.00,0.53,0.55}
\definecolor{turquoise}{rgb}{0.25,0.88,0.82}
\definecolor{violetred}{rgb}{0.82,0.13,0.56}
\definecolor{violet}{rgb}{0.93,0.51,0.93}
\definecolor{wheat1}{rgb}{1.00,0.91,0.73}
\definecolor{wheat2}{rgb}{0.93,0.85,0.68}
\definecolor{wheat3}{rgb}{0.80,0.73,0.59}
\definecolor{wheat4}{rgb}{0.55,0.49,0.40}
\definecolor{wheat}{rgb}{0.96,0.87,0.70}
\definecolor{whitesmoke}{rgb}{0.96,0.96,0.96}
\definecolor{white}{rgb}{1.00,1.00,1.00}
\definecolor{yellow1}{rgb}{1.00,1.00,0.00}
\definecolor{yellow2}{rgb}{0.93,0.93,0.00}
\definecolor{yellow3}{rgb}{0.80,0.80,0.00}
\definecolor{yellow4}{rgb}{0.55,0.55,0.00}
\definecolor{yellowgreen}{rgb}{0.60,0.80,0.20}
\definecolor{yellow}{rgb}{1.00,1.00,0.00}

\title[UV-derived recent star formation histories of nearby LIRGs]
    {A UV study of nearby luminous infrared galaxies: star formation histories and the role of AGN}

\author[S.~Kaviraj]
{S. Kaviraj$^{1}$\thanks{E-mail: skaviraj@astro.ox.ac.uk}\\
$^{1}$ Department of Physics, University of Oxford, Keble Road,
Oxford, OX1 3RH, UK}

\begin{document}
\maketitle

\label{firstpage}


\begin{abstract}
We employ UV and optical photometry, from the GALEX and SDSS
surveys respectively, to study the star formation histories (SFHs)
of 561 luminous infrared galaxies ($L_{IR}>10^{11}L_{\odot}$) in
the nearby ($z<0.2$) Universe. Visual inspection of a subsample of
galaxies with $r<16.8$ and $z<0.1$ (for which eyeball
classification of galaxy morphologies is reliable) indicates that
a small fraction ($\sim4\%$) have spheroidal or near-spheroidal
morphologies and could be progenitors of elliptical galaxies. The
remaining galaxies are morphologically late-type or ongoing
mergers. 61\% of the LIRGs do not show signs of interactions (at
the depth of the SDSS images), while the remaining objects are
either interacting ($\sim18\%$) or show post-merger morphologies
($\sim19\%$). {\color{black}Notwithstanding the high obscuration
in their stellar continua ($<A(FUV)> \sim 2.6$ mags, assuming a
Calzetti dust law), virtually all low-redshift LIRGs inhabit in
the UV `blue cloud'.} The (SSP-weighted) \emph{average} age of the
underlying stellar populations in these objects is typically 5-9
Gyrs, with a mean value of $\sim6.8$ Gyrs. $\sim60\%$ of the LIRG
population began their recent star formation (RSF) episode within
the last Gyr, while the remaining objects began their RSF episodes
1 to 3 Gyrs in the past. Up to 35\% of the stellar mass in the
remnant forms in these episodes - the mean value is $\sim15\%$.
The (decay) timescales of the star formation are typically
$\sim$few Gyrs, indicating that the star formation rate does not
decline significantly during the course of the burst. 14\% of the
LIRG population host (Type 2) AGN, {\color{black}with a hint that
the AGN fraction rises in interacting population (although low
number statistics hamper a robust result)}. The AGN hosts show UV
and optical colours that are redder than those of the normal
(non-AGN) population. There is no evidence for a systematically
higher dust content in the AGN hosts. AGN typically appear
$\sim0.5-0.7$ Gyrs after the onset of star formation and the
redder colours are a result of older RSF episodes, \emph{with no
measurable evidence of negative feedback from the AGN on the star
formation in their host galaxies.} Finally, we use the spheroidal
and near-spheroidal objects identified in this sample to study the
star formation that is plausibly induced by major mergers in the
low redshift Universe. The spheroidal remnants exhibit
(SSP-weighted) average ages of $\sim6.9$ Gyrs and form between 5
and 30\% of their stellar mass in the RSF episode, over time
periods between 0.3 and 4 Gyrs. We speculate that these galaxies
are the products of `mixed' major mergers, where at least one of
the progenitors has late-type morphology.
\end{abstract}




\section{Introduction}
{\color{black}The first infrared (IR) all-sky survey, performed by
the Infrared Astronomical Satellite (IRAS; Neugebauer et al.
1984), yielded a vast population of `Luminous Infrared Galaxies'
(LIRGs) that emit the bulk of their bolometric luminosity in the
far-infrared (FIR) wavelengths \citep{Sanders1996}. At
luminosities greater than $\sim10^{11}$L$_{\odot}$, they dominate
the galaxy census in the nearby Universe, becoming more numerous
(at comparable luminosities) than optically selected star-forming
galaxies, Seyferts and quasi-stellar objects (QSOs)
\citep[][]{Soifer1987,Soifer1991,Sanders1996}.} Traditionally, a
distinction is made between LIRGs, in which
$10^{11}L_{\odot}<L_{IR}<10^{12}L_{\odot}$ and a more powerful
class of `Ultra-Luminous Infrared Galaxies' (ULIRGs) which satisfy
$L_{IR}>10^{12}L_{\odot}$.

The infrared continua of LIRGs is dominated by thermal
re-radiation, by dust, of the ultra-violet (UV) flux from massive
main sequence stars in regions of intense star formation activity,
with a minor contribution from a dust-enshrouded AGN
\citep[e.g.][]{Rowanrobinson1989}. While the role of the AGN may
increase in significance for $L_{IR}>10^{12}L_{\odot}$, the
relative contributions from star formation and the AGN in ULIRGs
are still a matter of some debate \citep[see][and references
therein]{Joseph1999,Sanders1999,Elbaz2005,Teng2005}. While `cool'
ULIRGs ($f_{25}/f_{60}<0.25$) may simply involve scaled up
versions of the starbursts that drive LIRGs (e.g. Egami et al.
2004), a substantial AGN contribution to the infrared continuum is
expected in `warm' ULIRGS ($f_{25}/f_{60}>0.25$), which exhibit
optical, near-infrared and X-ray signatures of a powerful
(dust-enshrouded) AGN (e.g. Veilleux et al. 1999; see Chakrabarti
et al. 2007 for a theoretical interpretation).

\begin{figure*}
\begin{minipage}{172mm}
\begin{center}
$\begin{array}{cc}
\includegraphics[width=3.5in]{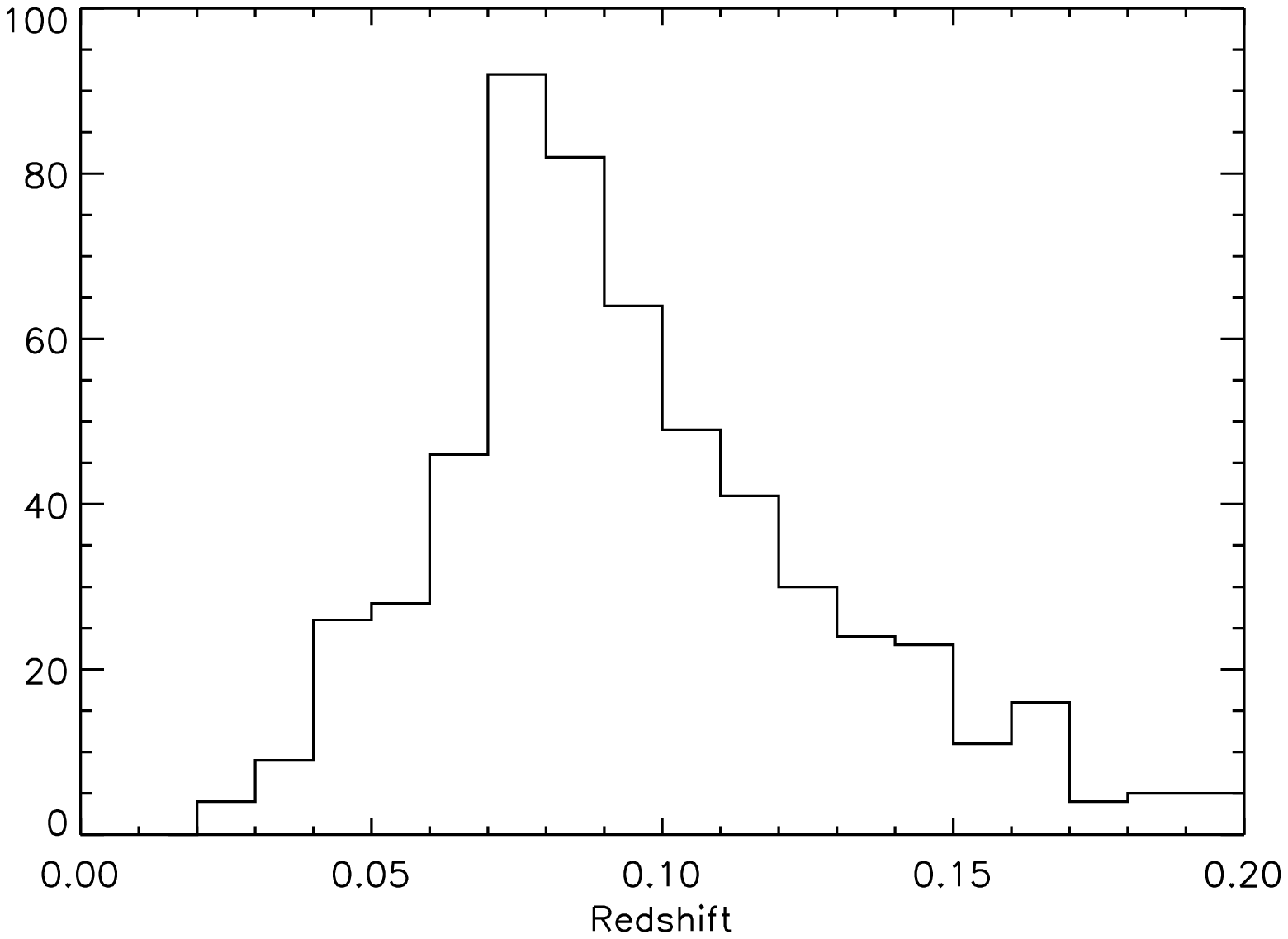}&\includegraphics[width=3.5in]{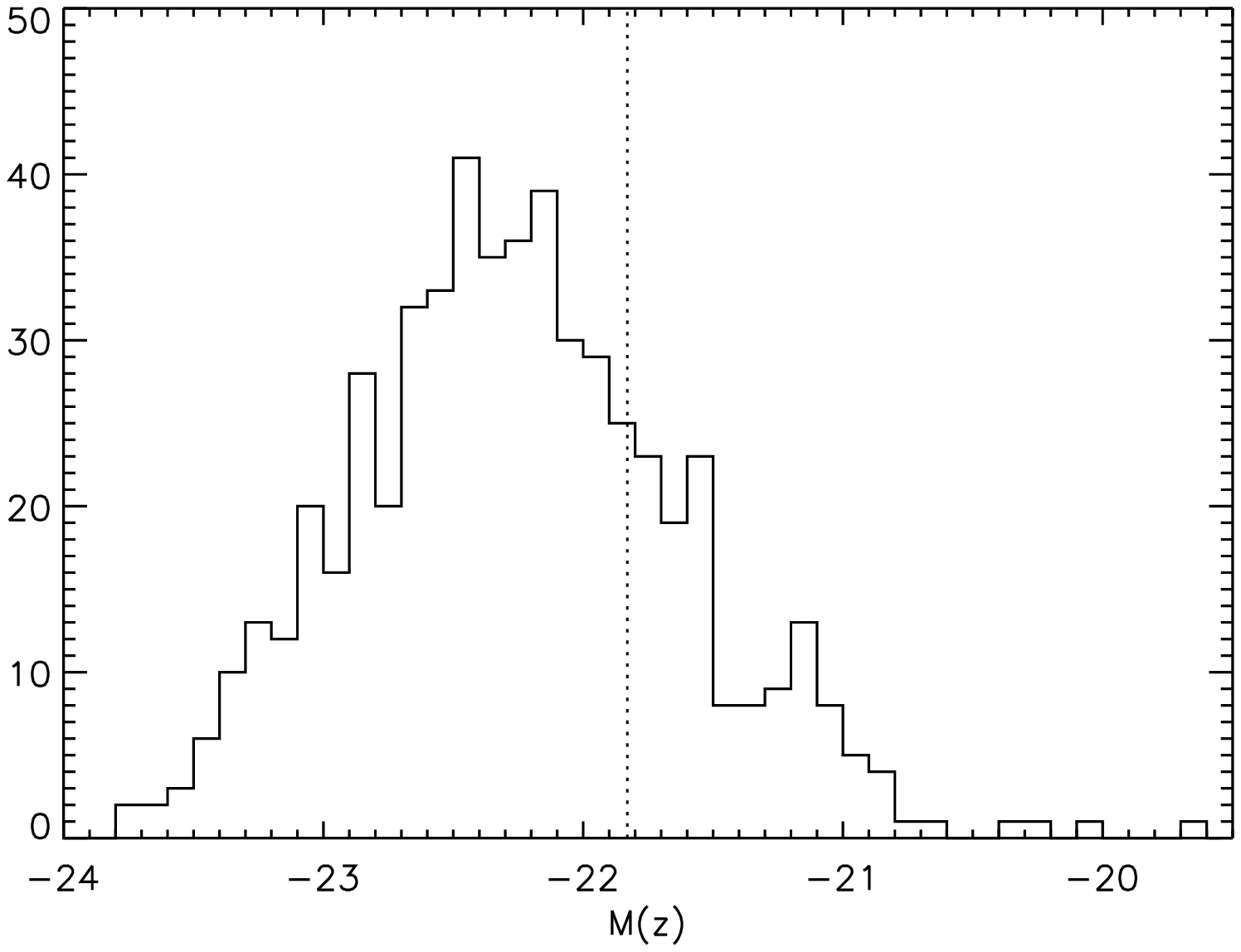}\\
\includegraphics[width=3.5in]{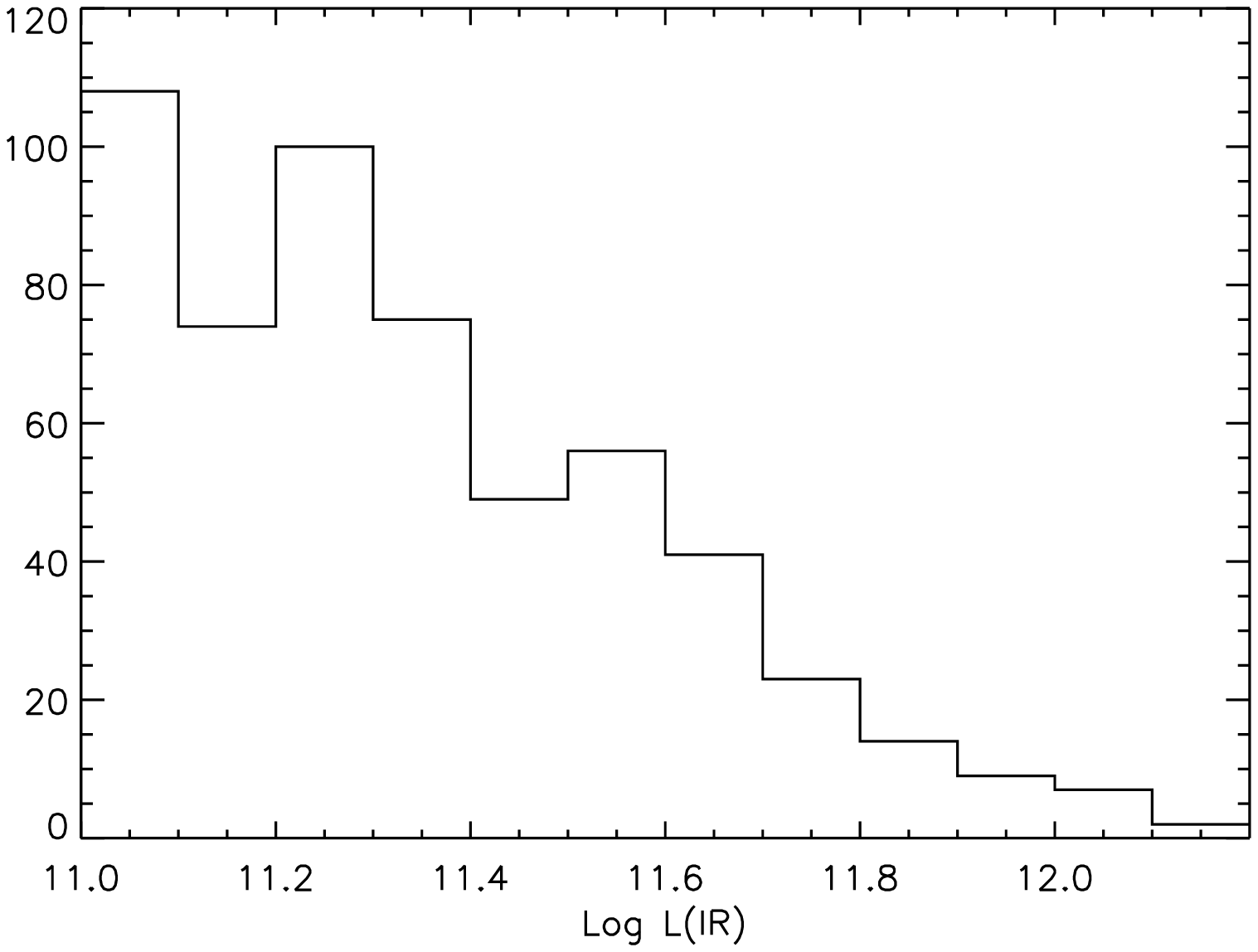}&\includegraphics[width=3.5in]{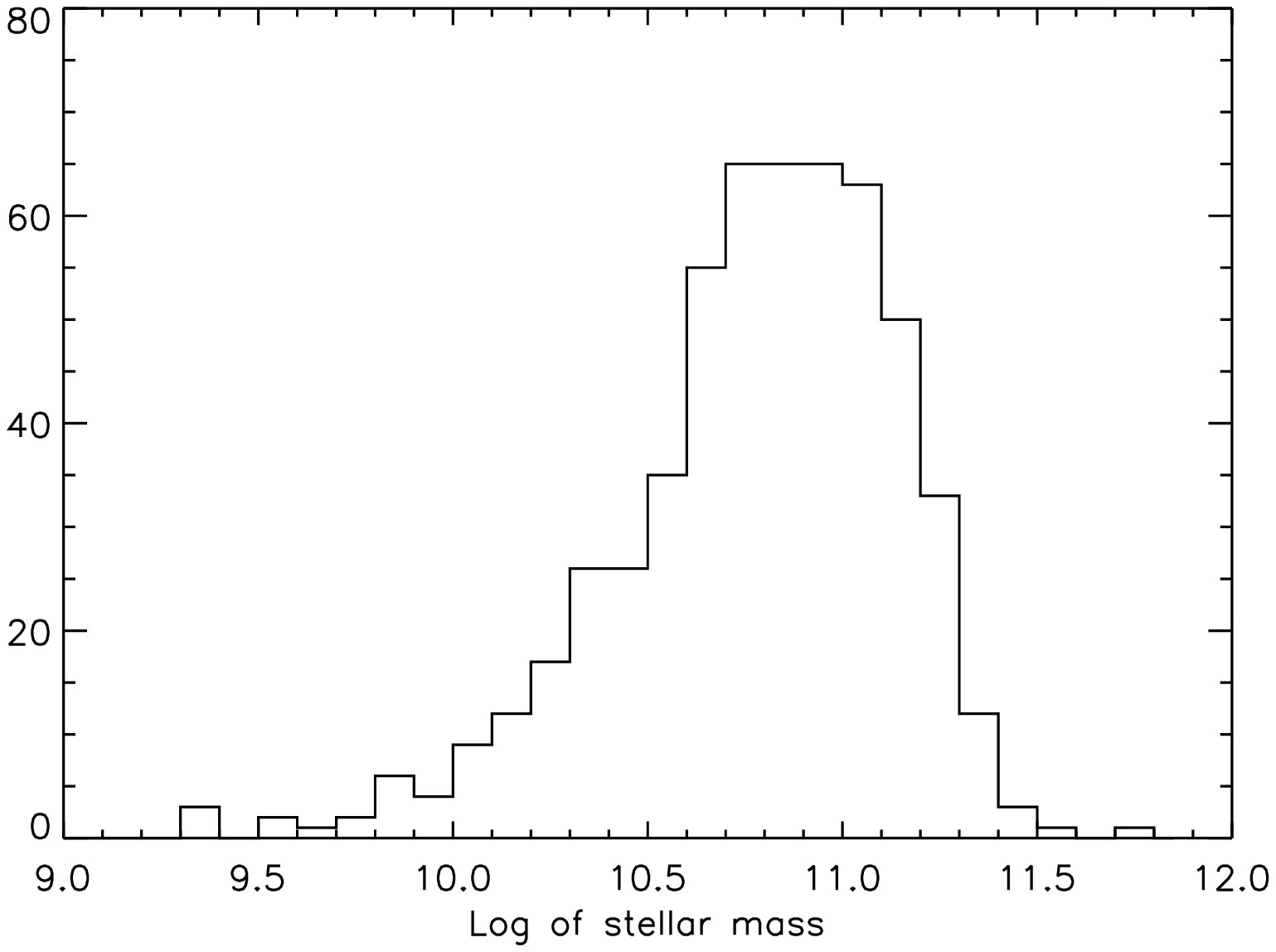}
\end{array}$
\caption{Basic properties of the LIRG sample studied in this
paper. TOP LEFT: Redshift distribution from the SDSS. TOP RIGHT:
Absolute z-band magnitudes. The value of M(z)$^*$ for early-type
galaxies (Bernardi et al. 2003) is indicated by the dotted red
line. Note that the $z$-band magnitudes are not K-corrected.
BOTTOM LEFT: Far-infrared luminosities from Cao et al. (2006) in
solar units. BOTTOM RIGHT: Stellar masses from Kauffmann et al.
(2003b) in solar units.} \label{fig:basic_properties}
\end{center}
\end{minipage}
\end{figure*}

The vigorous star formation activity itself is thought to be
induced either by (molecular) gas-rich mergers
\citep[e.g.][]{Sanders1996,Sanders2004,Laag2006} or by bar-driven
star formation in spiral galaxies \citep[e.g.][]{Wang2006}. While
the contributions to the LIRG population from mergers and spirals
at the high mass end ($M_*>10.5M_{\odot}$) are comparable in the
nearby Universe \citep{Wang2006}, ULIRGs are composed almost
exclusively of objects that appear to be in the advanced stages of
a major merger \citep[e.g. Arp 220;][]{Soifer1984,Sanders1996}.
The merger-driven nature of both LIRGs and ULIRGs is consistent
with the fact that both populations typically inhabit the field
\citep[e.g.][]{Goto2005,Zauderer2007}, where relative velocities
are lower and merging is more frequent.

{\color{black}The relative abundance of LIRGs shows a strong
evolution with redshift, consistent with the steep decline of the
cosmic star formation rate (SFR) that is observed over the last 8
billion years \citep[][]{Lilly1996,Madau1998}. While they are
relatively rare in the local Universe, comparison of the IR
luminosity functions from IRAS, at $z<0.3$, to surveys by the
Infrared Space Observatory (ISO; Taniguchi et al. 1997; Elbaz1999)
at intermediate redshifts ($0.3<z<1.5$) indicates that the
comoving space density of LIRGs rises by approximately three
orders of magnitude in the redshift range $0<z<1$
\citep{Sanders2003}. The large population of faint submillimeter
sources discovered using the SCUBA camera
\citep[e.g.][]{Smail1997} provides access to the ULIRG population
at $z>>1$ (with the ULIRGs dominating the energy density in this
redshift regime)}. These submillimeter studies \citep[see
e.g.][]{Lilly1999}, coupled with accurate determinations of the
mean redshift of the submillimeter population
\citep[e.g.][]{Chapman2003}, indicate that the space density of
ULIRGs at $z\sim2.4$ is a factor of $10^3$ higher than in the
local Universe. Furthermore, their total FIR luminosity density
exceeds the optical luminosity density from the general galaxy
population by up to an order of magnitude at $z>1.5$. Finally,
morphological studies of LIRGs over the last 8 billion years
indicate that the relative proportions of mergers and spirals
evolve with redshift \citep{Melbourne2005}, with the fraction of
mergers steadily increasing towards present-day
\citep[e.g.][]{Zheng2004,Bell2005,Wang2006}.

The advent of the GALEX ultra-violet (UV) space telescope (Martin
et al. 2005) is revolutionising our understanding of the local
Universe in the UV wavelengths (shortward of $\sim3000\AA$). The
UV spectrum is overwhelmingly dominated by flux from young massive
main sequence stars, making it an ideal probe of recent star
formation (RSF) in a variety of systems. Furthermore, the UV
remains largely unaffected by the age-metallicity degeneracy
\citep{Worthey1994} that typically plagues optical analyses
(Kaviraj et al. 2007a), making it an ideal photometric indicator
of RSF. Among the many successes of the new GALEX results has been
the detection of widespread low-level RSF in the nearby early-type
population (see Kaviraj et al. (2007b) for a review) and new
insights into the quenching mechanisms that operate in
post-starburst (E+A) galaxies at low redshift (Kaviraj et al.
2007c).

Given its sensitivity to recent and ongoing star formation, the
combination of UV and optical photometry is ideal for exploring
the star formation histories (SFHs) of the nearby LIRG population.
GALEX data has already been employed in the study of LIRGs and its
evolution with redshift. {\color{black}Burgarella et al. (2005a)
studied the GALEX photometry of 19 galaxies with spectroscopic
redshifts that are observed at $15\mu m$ as part of the European
Large-Area ISO Survey (ELAIS). The objects selected exhibit
infrared luminosities in the range
$10^{10}L_{\odot}<L_{IR}<10^{13}L_{\odot}$ and most are in the
nearby Universe ($z<0.3$). Comparison of their IR and UV
luminosities yields an average dust attenuation in the FUV
($\sim1530\AA$) of $A(FUV)\sim2.7$ mags (see also Burgarella et
al. 2005b)}. A recent study of 190 intermediate-redshift
($z\sim0.7$) LIRGs, observed with Spitzer/MIPS and GALEX
\citep{Buat2007}, reveals that the dust attenuation (as traced by
$L_{IR}/L_{UV}$) in these objects remains moderate to these
redshifts ($A_{FUV}<3$ mags), with a slight decrease in the mean
FUV dust attenuation of $\sim0.5$ mag in the redshift range
$0<z<0.7$, plausibly driven by lower metallicities and an increase
in the proportion of spirals that are LIRGs (under the assumption
that merging systems are more affected by dust).

\begin{figure}
\begin{center}
$\begin{array}{c}
\includegraphics[width=3in]{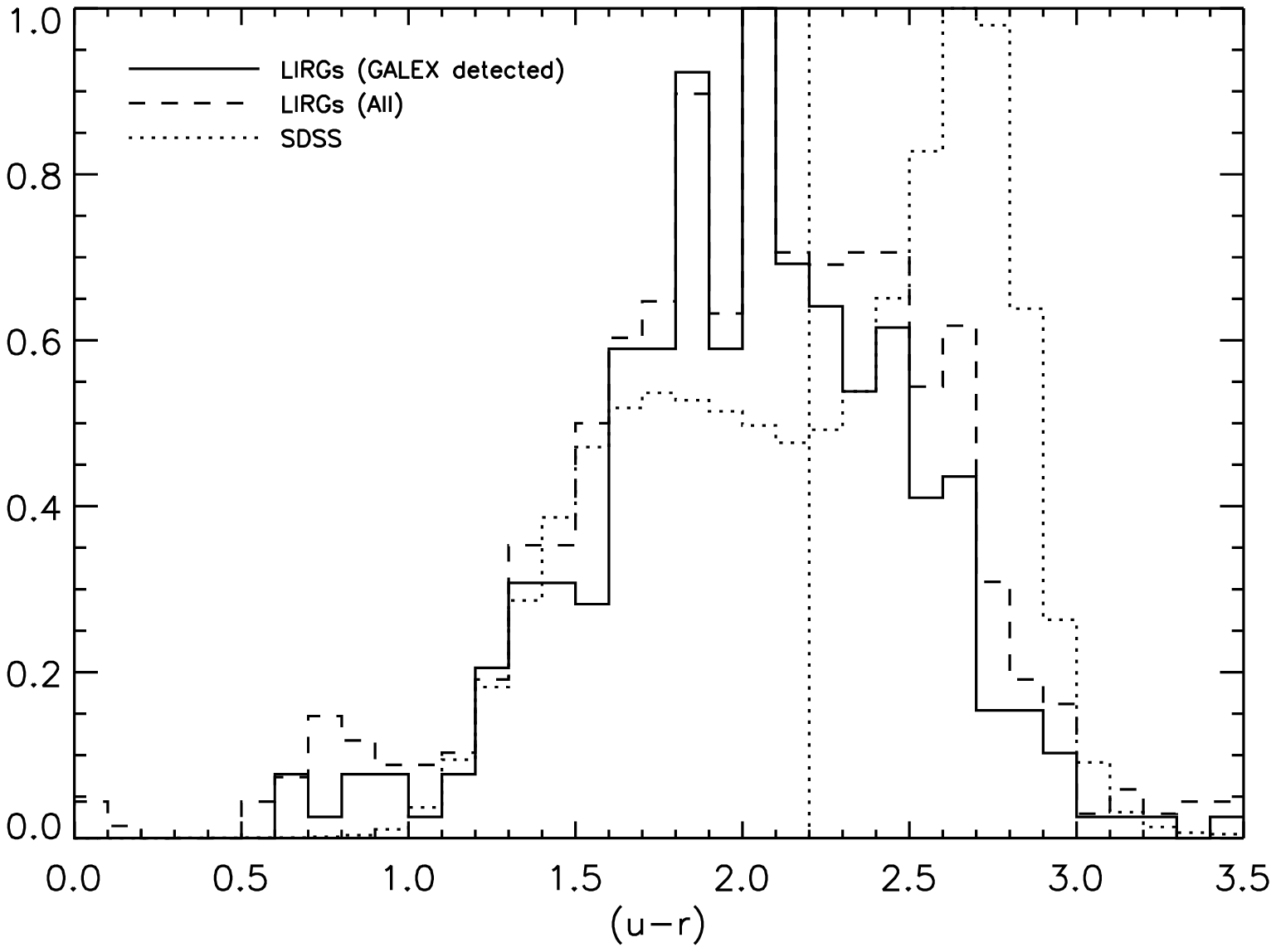}\\
\includegraphics[width=3in]{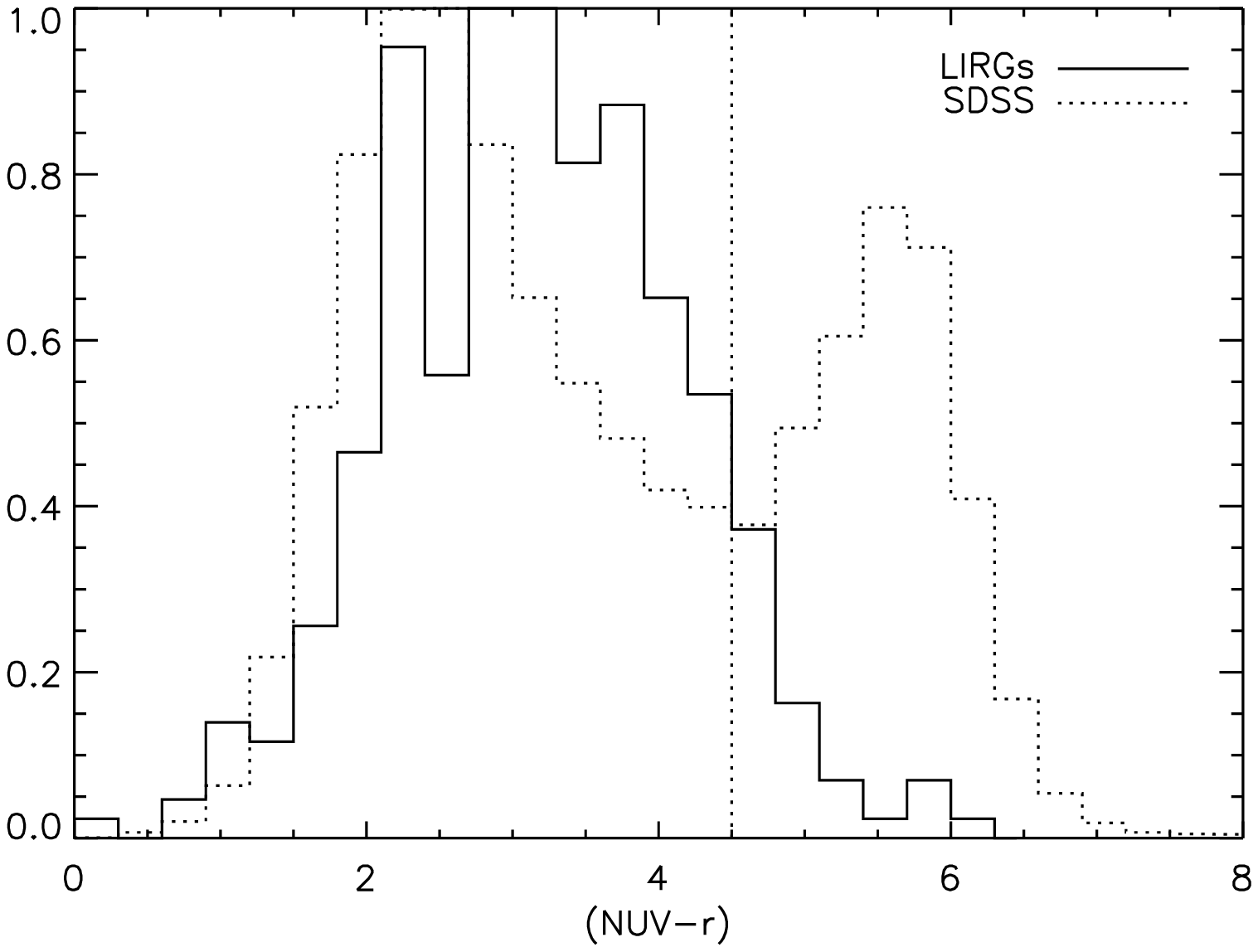}\\
\end{array}$
\caption{{\color{black}Comparison of the $(u-r)$ (top) and
$(NUV-r)$ (bottom) colours of the LIRG population (solid lines) to
the general galaxy population drawn from the SDSS (dotted lines).
In both plots the separation between the blue and red peaks is
indicated using a vertical dotted line. The comparisons are
restricted to a volume limited sample ($z<0.1$, $r<17.77$). Note
that while the relative proportions of galaxies in the red and
blue peaks change depending on the magnitude cut employed, the
colour at which the population divides remains stable (see e.g.
Figure 2 in Strateva et al. 2001).} {\color{black}For the $(u-r)$
comparison, we show histograms for both the LIRG population that
has been detected by GALEX (solid line) and the full LIRG sample
drawn from the SDSS (dashed line).}} \label{fig:lirgs_comparison}
\end{center}
\end{figure}

In this paper, we perform a quantitative study of the SFHs of a
large sample of LIRGs in the nearby Universe ($z<0.2$), by
combining their UV (GALEX) and optical (SDSS; Adelman-McCarthy et
al. 2004) photometry. This work utilises many of the theoretical
tools that have been developed and successfully applied to
GALEX-SDSS data from various galaxy populations, in particular
early-types and E+A systems. The novelty of this study is (a) its
quantitative nature - we quantify the SFHs of an unprecedentedly
large sample of nearby LIRGs (b) the incorporation of the UV into
the parameter estimation, which provides robust constraints on the
recent SFH and (c) that we explore the role of AGN in these types
of objects and search for signatures of feedback from the AGN on
the SFH and gas content of their host galaxies.

In Section 2, we present the sample of galaxies used in this study
and describe the process of visual inspection used to classify
galaxies according to their morphology and interaction status.
Section 3 describes the theoretical machinery used to quantify the
SFH of each LIRG object in our sample. In Section 4 we explore the
SFHs of the LIRG population as a function of both their
morphologies (early/late/mergers) and their interaction status
(isolated/interacting/post-mergers), while in Section 5 we study
the onset of AGN in these systems and explore whether there is
plausible evidence for AGN feedback in these galaxies. Finally, in
Section 6 we use the derived SFH parameters of spheroidal objects
identified in this sample to explore the characteristics of major
mergers in the nearby Universe.


\section{The sample}
\subsection{Photometric data and basic properties}
Our study is based on a  catalog of LIRGs, drawn from the SDSS
Data Release 2 (DR2), compiled by \citet{Cao2006} by
cross-correlating the IRAS Faint Source Catalog \citep{Moshir1992}
with the SDSS DR2. This catalog has been cross-matched with
publicly available $UV$ photometry from the third data release of
the GALEX mission in its {\color{black}Medium Imaging Survey (MIS)
mode (exposure time $\sim$ 1ks)}. We restrict ourselves to objects
below a redshift of 0.2, since outside this range the GALEX
filters do not trace the rest-frame UV. GALEX provides two $UV$
filters: the far-ultraviolet ($FUV$), centred at $\sim1530\AA$ and
the near-ultraviolet ($NUV$), centred at $\sim2310\AA$. This
cross-matching produces 561 LIRGs with \emph{at least} an MIS
detection in the $NUV$ filter, of which 413 have detections in
both the $FUV$ and the $NUV$. {\color{black}Note that we have
removed Type I AGN from the sample because they could potentially
contaminate the UV spectrum and skew our parameter estimation.}

{\color{black} The final sample studied in this paper corresponds
to the following limits: $0.02<z<0.2$, $m_r<17.77$,
$16.23<m_{NUV}<22.87$, $15.86<m_{FUV}<23.2$. The spectroscopic
confidence ($z_{conf}$) is required to be better than 0.7 for each
object. The redshift and magnitude limits of this study are
similar to those of Wyder et al. (2007), who have performed a
comprehensive analysis of the UV-optical colours of galaxies that
have been detected in both the SDSS and GALEX surveys. Note
however that, unlike this study, Wyder et al. (2007) also imposed
a bright $r$-band cut of 14.5 mags, because the incompleteness of
the SDSS spectroscopic coverage increases beyond this limit due to
incomplete deblending of extended objects \citep[see
e.g.][]{Strauss2002}. We choose not to impose this cut here as
only 1.3\% of the sample studied in this paper has $r<14.5$ and
visual inspection of each object confirms that there are no
anomalies in the SDSS pipeline processing.}

In Figure \ref{fig:basic_properties} we present some basic
properties of the LIRG sample used in this study - the redshift
distribution, absolute $z$-band magnitudes, FIR luminosities and
stellar masses. {\color{black}In Figure \ref{fig:lirgs_comparison}
we compare the $(u-r)$ and $(NUV-r)$ colours of the LIRG
population (solid and dashed lines) to the general galaxy
population drawn from the SDSS (dotted lines). {\color{black}Note
that we show $(u-r)$ histograms for both the LIRG population that
has been detected by GALEX (solid line) and the full LIRG sample
drawn from the SDSS (dashed line).} In both plots the separation
between the blue and red peaks in the general galaxy population is
indicated using a vertical dotted line. The comparisons are
restricted to a volume limited sample in the SDSS at $z=0.1$.
While the relative proportions of general galaxies in the red and
blue peaks may change depending on the magnitude cut employed, the
colour at which the population divides remains stable (see e.g.
Figure 2 in Strateva et al. 2001).

We find that, in the $(u-r)$ colour, the LIRG population peaks
around the colour bimodality point ($u-r\sim2.2$) with a large
fraction of the LIRGs falling in the blue cloud. In this optical
colour the LIRG population seems to largely occupy intermediate
colours between the blue cloud and red sequence. The situation in
the $(NUV-r)$ colour is somewhat different. The LIRGs peak in the
UV blue cloud, with a minor tail into the red sequence. The plots
shown here are a useful benchmark for similar comparisons at high
redshift. Both dust obscuration and the strength of star formation
activity is likely to evolve with increasing look-back time and
similar plots from high-redshift surveys are keenly anticipated.}

{\color{black}Finally, we note some points regarding the SDSS and
GALEX data that underpin this study. A potential concern for
objects, typically for galaxies at low redshift, is pipeline
`shredding' of extended objects into smaller sub-components. In
the very local Universe (redshifts lower than z=0.01), especially
in large galaxies that have extremely irregular light
distributions, shredding is a potential concern, particularly for
GALEX detections in the shallowest All-sky Imaging Survey (AIS)
mode (exposure time $\sim$100s) because the data is more noisy
(Wyder et al. 2007, Ted Wyder, Mark Seibert priv comm.). For
galaxies detected in the GALEX MIS mode, such as those used in
this study, shredding is not an issue in our target redshift range
($0.02<z<0.2$). To ensure the quality of the GALEX data and the
cross-matching to SDSS, a random 30\% of the GALEX images were
visually inspected to verify that (a) there was no evidence for
shredding and (b) that cross-matching to the corresponding SDSS
objects was robust. In addition, as a final check against
deblending issues, we have checked that there is no systematic
evolution in the average value or scatter of the $(NUV-r)$ colours
of the LIRGs population, which would be expected if there was
systematically higher shredding at lower redshifts.

Note that since each galaxy in this study is assigned a morphology
and interaction status through visual inspection of its SDSS image
(see next section), we can robustly verify a lack of shredding in
the entire SDSS dataset used here. Finally, a minor fraction of
GALEX objects ($\sim20$\%, see e.g. Seibert et al. 2005) have
multiple SDSS matches, due to detections in more than one GALEX
field. For such objects, we select the GALEX detection with the
highest exposure time for the subsequent analysis.}

\begin{figure*}
\begin{minipage}{172mm}
\begin{center}
\includegraphics[height=0.5\textheight]{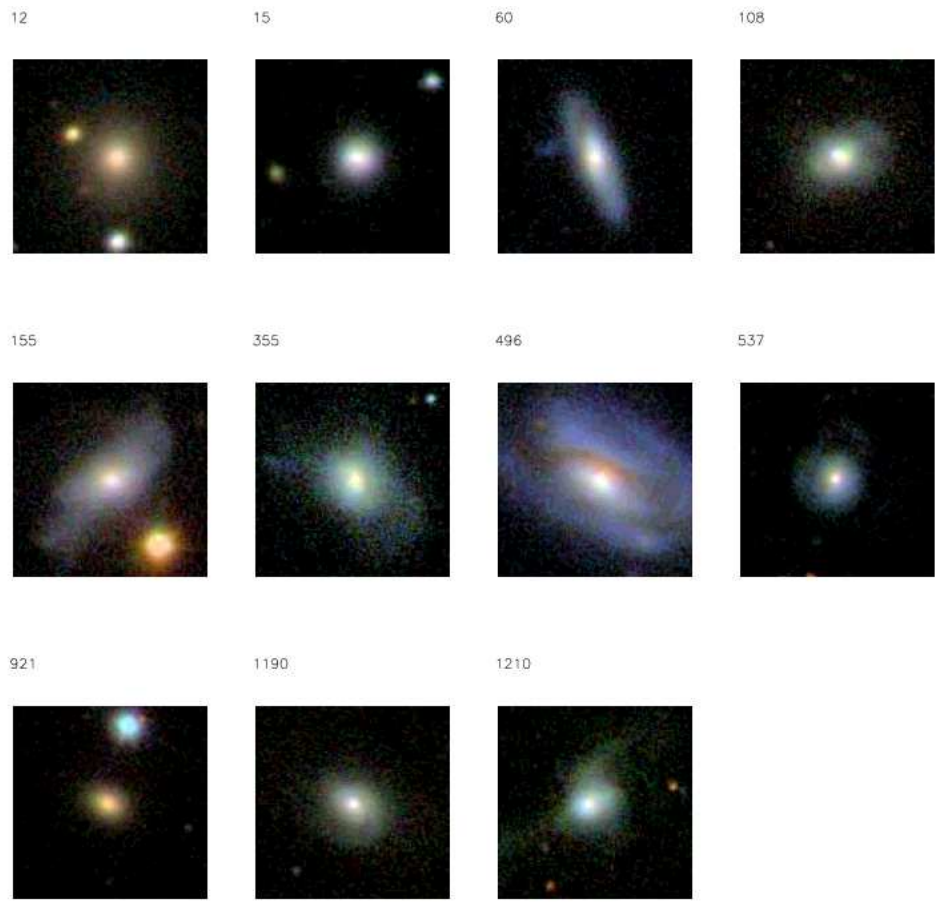}
\caption{`Early-type' galaxies identified in the LIRG population
through visual inspection of the subsample of LIRGs that satisfy
$r<16.8$ and $z<0.1$ (see Section 2.2 for the definition of
early-type used in this study). Note that these objects typically
show near-spheroidal morphology in their cores but exhibit
morphological disturbances consistent with recent merging (108,
155, 355, 496, 537, 1190 and 1210). Object 496 is plausibly an
`NGC 5128 in the making'. Galaxy 60 appears to be in the final
stages of accreting a small companion. While this object could be
an inclined disk rather than a spheroid, it is most likely to be
an S0 system with a dust feature in the northern part of the
galaxy. The dust features appears to be spatially coincident with
the accreted companion. Only three objects in this subset (12, 15
and 921) appear relaxed.} \label{fig:etg_gallery}
\end{center}
\end{minipage}
\end{figure*}

\begin{figure*}
\begin{minipage}{172mm}
\begin{center}
\includegraphics[height=0.5\textheight]{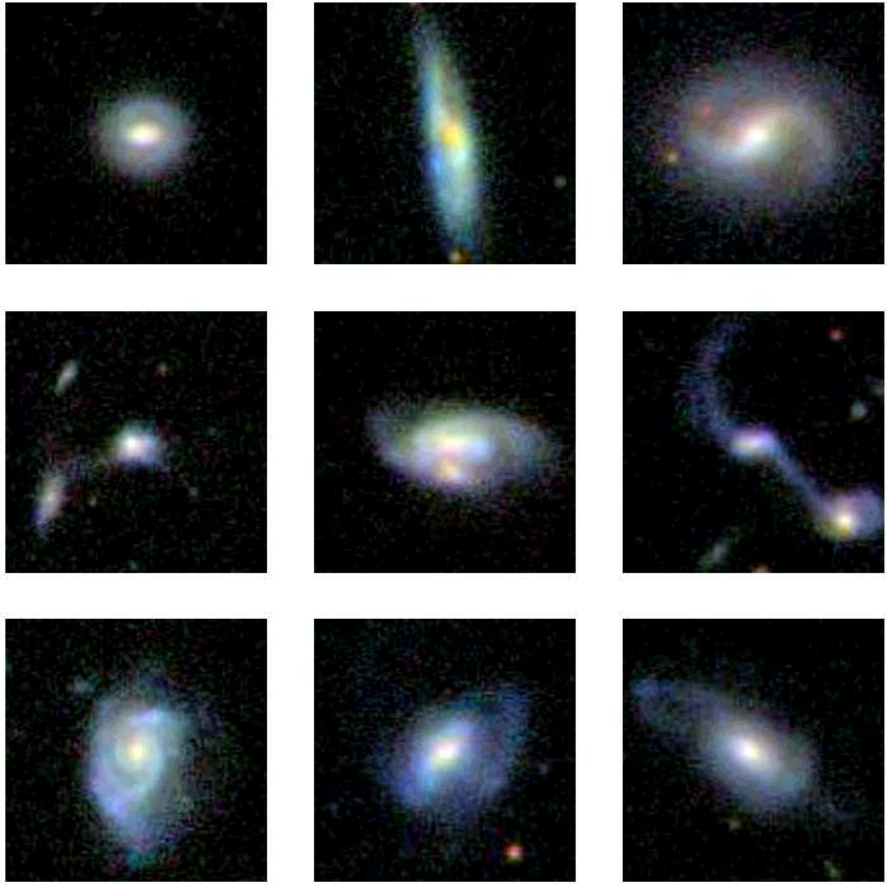}
\caption{Examples of `late-type' galaxies in the LIRG sample. The
top row shows late-types classified as `isolated'. The middle row
shows late-types identified as `interacting/merging', while the
bottom row indicates late-types that are classified as
`post-mergers'.} \label{fig:spirals_gallery}
\end{center}
\end{minipage}
\end{figure*}


\subsection{Visual inspection}

\begin{table*}
\begin{minipage}{150mm}
\begin{center}
$\begin{array}{cc}
\includegraphics[width=5in]{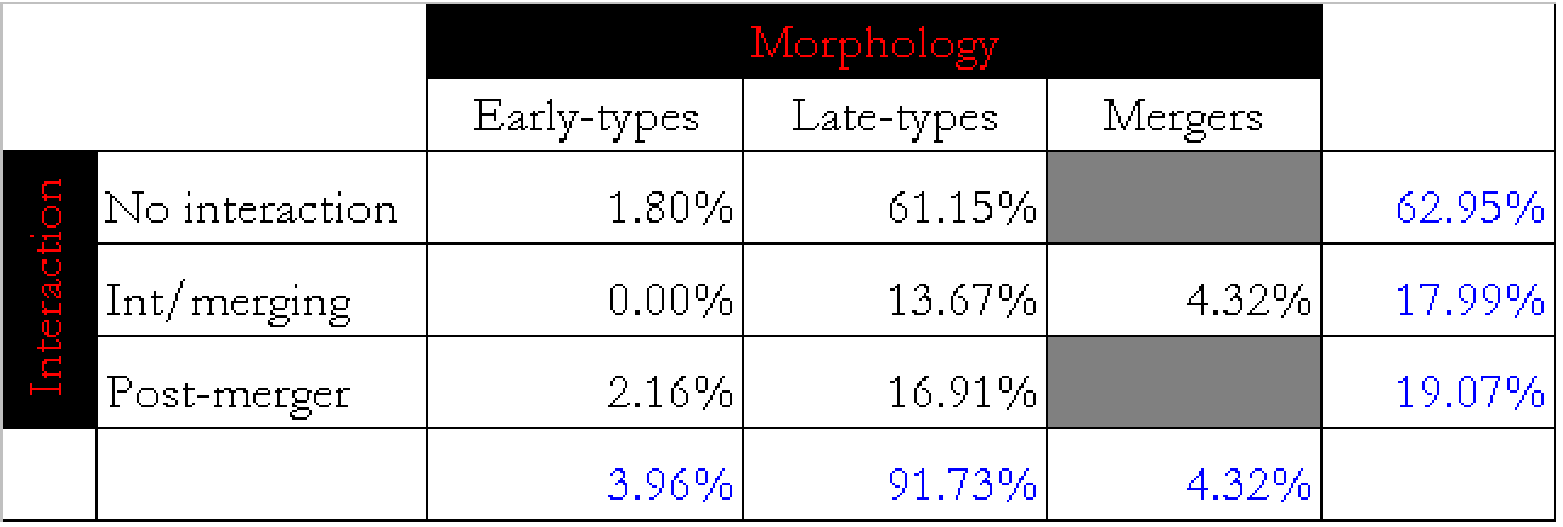}&
\end{array}$
\caption{The LIRG sample split by morphology and interaction
status (see Section 2 for descriptions of the categories). The
totals for each category are shown in blue on the right and bottom
ends of the table. Note that ongoing mergers are automatically
labelled as interacting/merging. The sample is restricted to
$r<16.8$ and $z<0.1$ since eyeball inspection from SDSS images is
reliable within these magnitude and redshift ranges (see Section 2
in Kaviraj et al. 2007b).} \label{fig:morphintgrid}
\end{center}
\end{minipage}
\end{table*}

\begin{figure*}
\begin{minipage}{172mm}
\begin{center}
\includegraphics[height=0.5\textheight]{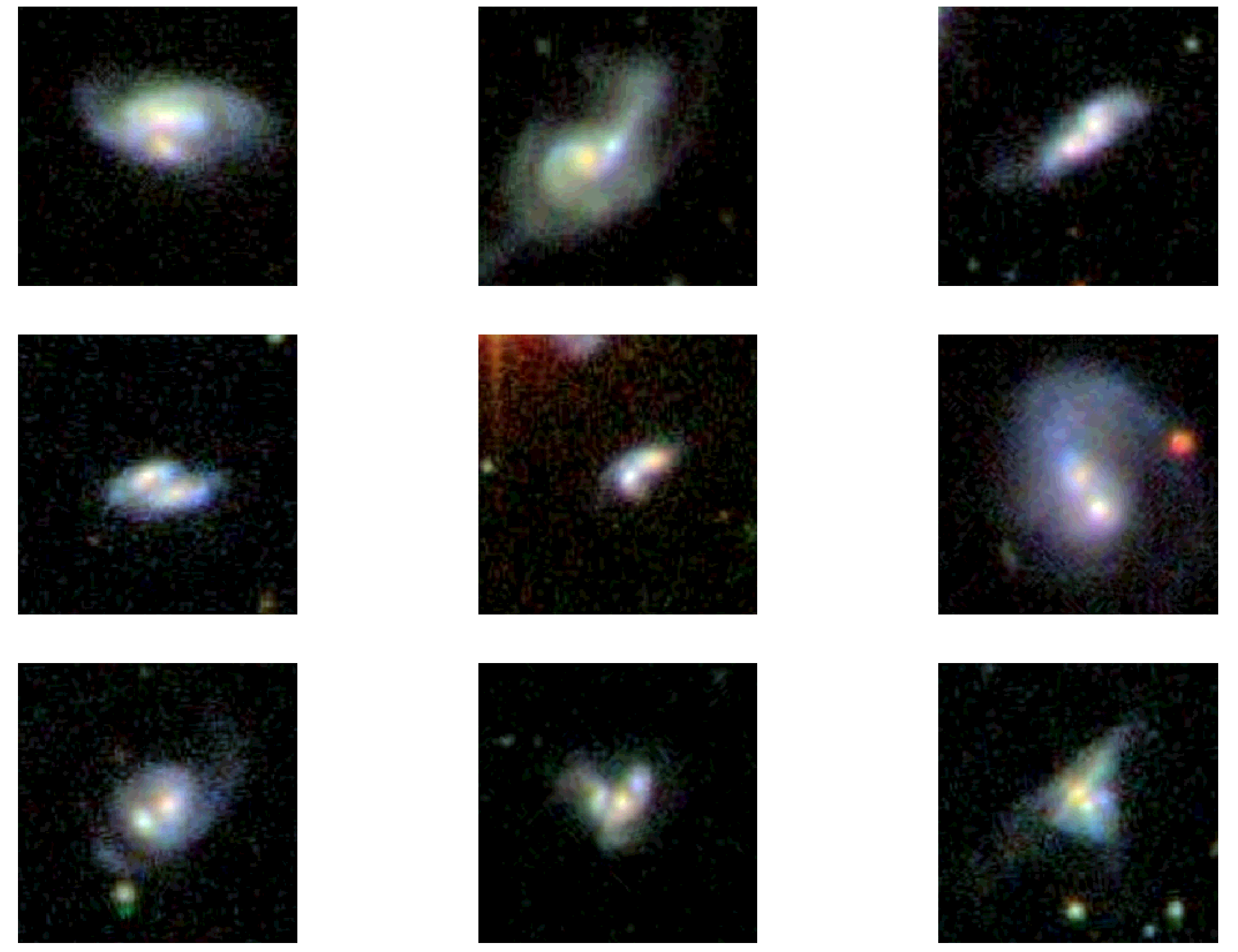}
\caption{Examples of systems labelled as `mergers' in the LIRG
sample.} \label{fig:mergers_gallery}
\end{center}
\end{minipage}
\end{figure*}

A basic aim of this work is to correlate the SFHs of the LIRG
population with their morphologies and `interaction status'. We
divide galaxies into three morphologies: `early-types' (spheroidal
or near-spheroidal objects that are likely to be the progenitors
of elliptical galaxies), `late-types' and `mergers' (where
\emph{two} separate objects are involved and are already partially
fused). The interaction status assigned to each object also falls
into one of three categories: `isolated', `interacting/merging' or
`post-merger'. Systems that are morphologically identified as
`mergers' are automatically labelled as being
`interacting/merging'.

While it becomes more difficult to decipher spiral features in
higher redshift images, prior experience with early-type galaxies
suggests that galaxy morphologies can be determined from SDSS
images with good precision out to $z\sim0.1$ if the objects have
$r\lesssim16.8$ (see Section 2 in Kaviraj et al. 2007b). In
particular, if morphological classifications are kept fairly
coarse (e.g. early/late), as is the case in this study, then
tell-tale signs of disks can be successfully identified in the
\emph{composite} images, since disk-like features readily appear
in the bluer optical filters (e.g. $g$-band).

We therefore begin by visually classifying all the objects in this
sample that satisfy the fiducial criteria of $z<0.1$ and $r<16.8$.
We present the results of this visual inspection in Table
\ref{fig:morphintgrid}. We note first that, in agreement with
previous studies of the LIRG population, the fraction of
early-types is almost negligible. In Figure \ref{fig:etg_gallery}
we present \emph{all} galaxies labelled as early-type in this
visual inspection. Recall, however, that our definition of
`early-type' includes \emph{near-spheroidal} objects which are
likely to be progenitors of elliptical galaxies. It is apparent
that many of galaxies in Figure \ref{fig:etg_gallery}, while
having spheroidal cores, are not completely relaxed. Tidal
features and evidence of accretion are widespread in the galaxy
images indicating that they are merger remnants, although a few
systems do exist that have relaxed morphologies, similar to the
star-forming ellipticals discovered by \citet{Fukugita2004}.
Relaxed early-types with high LIRG-like SFRs are likely to be very
rare in the general galaxy population and their detection clearly
depends on having a large sample of objects, as is the case in
this study.

Since the early-type fraction is small, we are effectively left
with two morphologies: mergers and late-types. In essence, this
reduces to simply identifying the mergers, since the absence of
early-types implies that all `non-mergers' are late-types.
Fortunately, given their distinct morphologies (typically double
cores with asymmetric tidal distortions), mergers can readily be
identified even when they are fainter than $r\sim16.8$ or further
away than $z\sim0.1$. Since it is reasonable to assume that the
fraction of early-types remains small across the entire redshift
range of this study ($z<0.2$), this also implies that we can
extract a reasonably robust morphological classification, even in
the subsample of galaxies that is fainter than $r\sim16.8$ and
further away than $z\sim0.1$. We stress, however, that this
\emph{only} works for this particular sample because (a) our
classification criteria are sufficiently coarse
(early/late/mergers) (b) the early-type fraction is very small and
(c) mergers can be readily identified even if the systems are
relatively faint or distant. We also note that the interaction
status of a galaxy can be determined accurately, even if the
morphology of the object is uncertain. Identification of
interacting/merging galaxies is straightforward, given the
presence of an interacting neighbour and tidal bridges (which are
readily visible in SDSS images of the nearby Universe).
Post-mergers, i.e. systems where a single dominant central object
shows asymmetric tidal features, are also readily identified.

Based on these arguments we visually classify all galaxies in
terms their morphologies (early/late/mergers) and interaction
status (isolated/interacting/post-mergers) in the redshift range
$z<0.15$. The redshift limit is chosen because (a) it encloses
most of the galaxies in the sample (see Figure
\ref{fig:basic_properties}) and (b) because tidal distortions in
post-mergers at the limit of this redshift range already appear
quite faint and it might be difficult to detect such post-merger
signatures in galaxies much further away. Note that we \emph{only}
attempt to identify early-types in the subsample of galaxies where
$r<16.8$ and $z<0.1$. Therefore, the morphological classification
for the complete sample of galaxies, where $z<0.15$, is likely to
be incorrect by $\sim4\%$ (the early-type fraction found for
$r<16.8$ and $z<0.1$). However, given that this is very small, it
does not affect our subsequent analysis very significantly. Figure
\ref{fig:spirals_gallery} shows examples of objects labelled as
late-type. The top row indicates late-types classified as
`isolated'. The middle row shows late-types identified as
`interacting/merging', while the bottom row shows late-types that
are `post-mergers'. Finally, in Figure \ref{fig:mergers_gallery}
we present examples of objects labelled as mergers.

We note that the visual classifications performed in this study
are clearly limited by the depth of the SDSS images, which involve
50s exposures on a 2.5m telescope. However, the detection of faint
or red features may require deeper imaging. This is demonstrated
by the recent work of \citet{VD2005} who found widespread (and
perhaps unexpected) post-merger signatures in a large fraction
($>70\%$) of red early-type galaxies using $\sim$27,000 second
exposures on 4m class telescopes. These tidal features extend to
spatial scales $>50$ kpc and are, almost without exception,
undetectable in their shallower SDSS images. It is possible,
therefore, that an unknown fraction of the LIRGs labelled as
`isolated' in this study may actually be post-mergers with very
faint tidal features. While the tidal features in LIRGs are likely
not to be red, indicating that this problem may not be as acute as
for the red early-types in the \citet{VD2005} study, it is likely
that the fraction of LIRGs classified as isolated has been
overestimated coupled with a corresponding underestimate of the
LIRG population that are classified as post-mergers.


\subsection{Stellar masses and AGN diagnostics}
In addition to the photometric and visual inspection data, we use
the publicly available Garching SDSS
catalog\footnote{http://www.mpa-garching.mpg.de/SDSS/DR4/} to add
stellar masses and AGN diagnostics for each galaxy in our catalog.
Type 2 AGN can be identified using combinations of optical
emission line ratios (see e.g. Baldwin et al. 1981; Kauffmann et
al. 2003a; Brinchmann et al. 2004). Here, we use AGN
classifications derived by \citet{Brinchmann2004} from line index
measurements calculated using the code of \citet{Tremonti2004}.
Galaxies fall into one of three categories: AGN, composites (which
show signatures of both AGN and star formation) and normal (where
the emission lines are dominated by star formation or where the
galaxy does not have the requisite S/N in the emission lines to
perform a classification in the first place). Note that Type 1 AGN
are removed from this analysis since their UV spectra is likely to
be dominated by the active nucleus which, in turn, will perturb
the parameter estimation used to quantify the SFHs. The stellar
masses are taken from the catalog of Kauffmann et al. (2003b).


\section{Parameter estimation}
{\color{black}We estimate parameters governing the star formation
history (SFH) of each LIRG in this sample by comparing its $(FUV,
NUV, u, g, r, i, z)$ photometry to a library of synthetic
photometry which is constructed using a large collection of model
SFHs as follows}.

Each {\color{black}model} SFH is constructed by modelling the
underlying population of the galaxy by a `primary burst' (PB)
which is assumed to be instantaneous, and the recent period of
star formation (which is what we are primarily interested in)
using an exponential burst. We refer to this recent episode as the
`secondary burst'. In the notation used below, we denote the
primary burst using the subscript `1' and the secondary burst by
the index `2'.

\begin{figure}
\begin{center}
\includegraphics[width=3.5in]{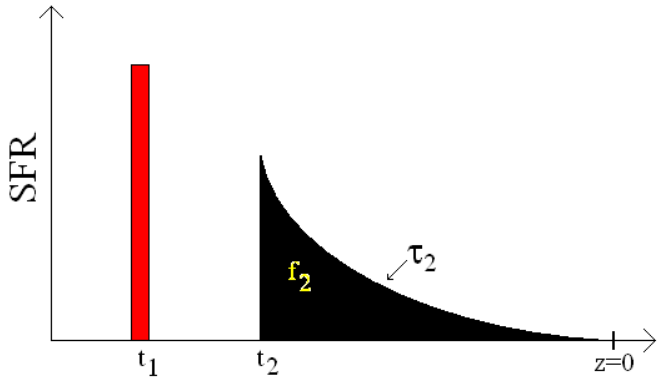}
\caption{Schematic representation of model star formation
histories (see Section 3 for details). $t_1$ is the primary burst
(PB) age. $t_2$ is the secondary burst (SB) age. $f_2$ is the SB
mass fraction and $\tau_2$ is the SB timescale.}
\label{fig:sfh_cartoon}
\end{center}
\end{figure}

Figure \ref{fig:sfh_cartoon} shows a schematic representation of
the model SFHs. The principal free parameters in this analysis are
the age of the primary burst ($t_1$), and the age ($t_2$), mass
fraction ($f_2$) and timescale ($\tau_2$) of the secondary burst.
$t_1$ is allowed to vary from 1 Gyrs to 12 Gyrs, while $t_2$ is
allowed to vary from 0.05 Gyrs to 12 Gyrs in the rest-frame of the
galaxy. $f_2$ varies between 0 and 1 and $\tau_2$ is allowed to
vary from 0.01 Gyrs to 9 Gyrs. Since there are only two bursts,
$f_1=1-f_2$.

Since it is modelled as being instantaneous, the PB age ($t_1$)
reflects the (SSP-weighted) \emph{average} age of the underlying
stellar population in the galaxy. As our subsequent analysis
indicates, the values of $t_1$ are typically higher than 5 Gyrs,
so that the primary burst does not affect the UV colours of the
galaxy (which are dominated by stars less than 2 Gyrs old - see
Figure 7 of Kaviraj et al. 2007b).

To build the library of synthetic photometry, each model SFH is
combined with a single metallicity in the range 0.1Z$_{\odot}$ to
2.5Z$_{\odot}$ and a value of dust extinction parametrised by
$E_{B-V}$ in the range 0 to 1. {\color{black}Photometric
predictions for each model SFH are generated by combining it with
the chosen metallicity and $E_{B-V}$ values and convolving with
the stellar models of \citet{Yi2003} through the GALEX $FUV$,
$NUV$ and SDSS $u, g, r, i, z$ filters. Note that we use the
empirical dust prescriptions of \citet{Calzetti2000} to compute
the dust-extincted SEDs. This procedure yields a synthetic library
of $\sim1.8$ million models.}

Since our sample is comprised of objects across a range in
redshift, equivalent libraries are constructed at redshift
intervals of $\delta z=0.01$. A fine redshift grid is essential in
such a low redshift study because a small change in redshift
produces a relatively large change in look-back time over which
the $UV$ flux can change substantially, inducing
`K-correction-like' errors into the analysis.

\begin{figure}
\begin{center}
$\begin{array}{cc}
\includegraphics[width=3.5in]{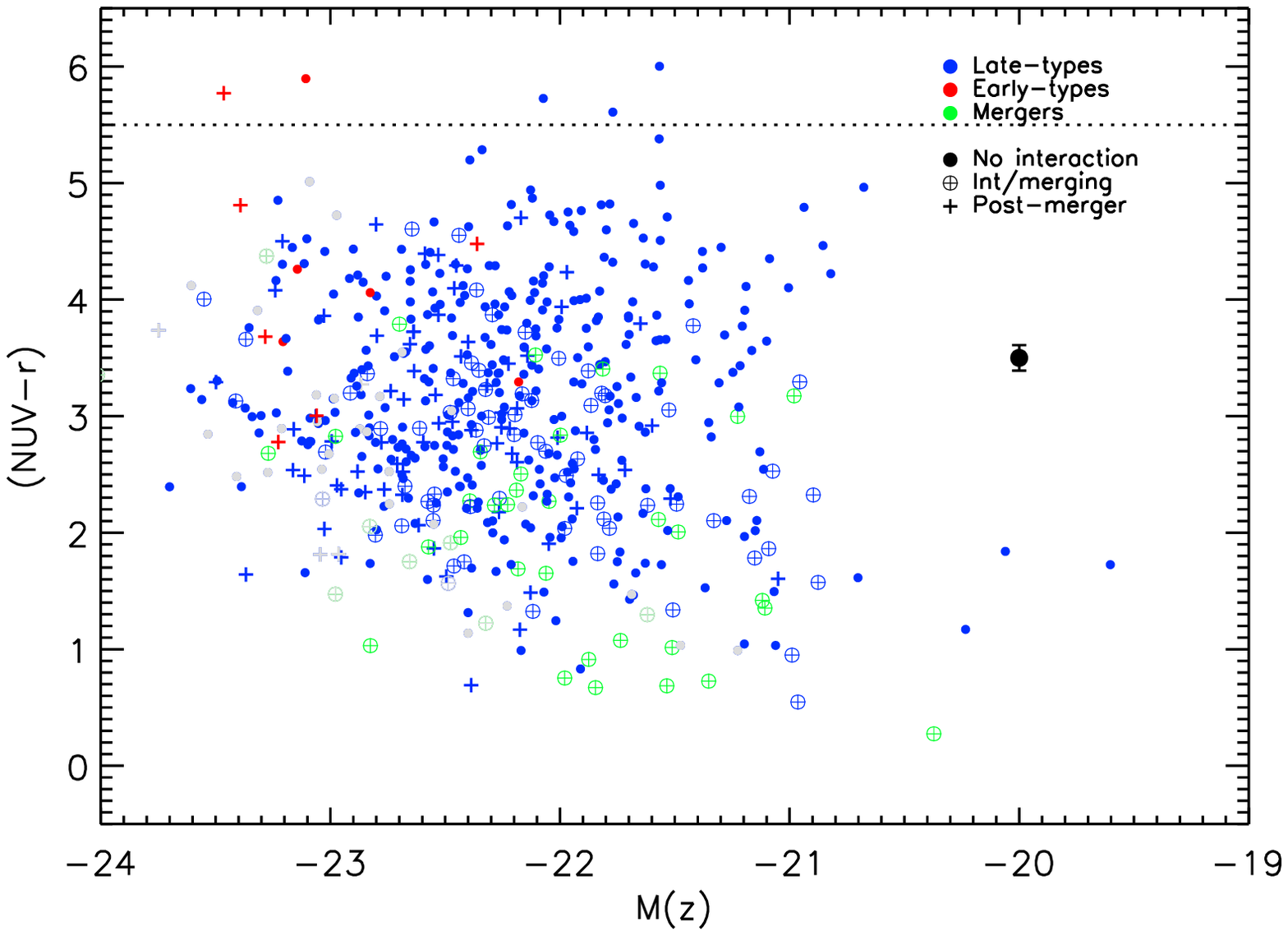}\\
\includegraphics[width=3.5in]{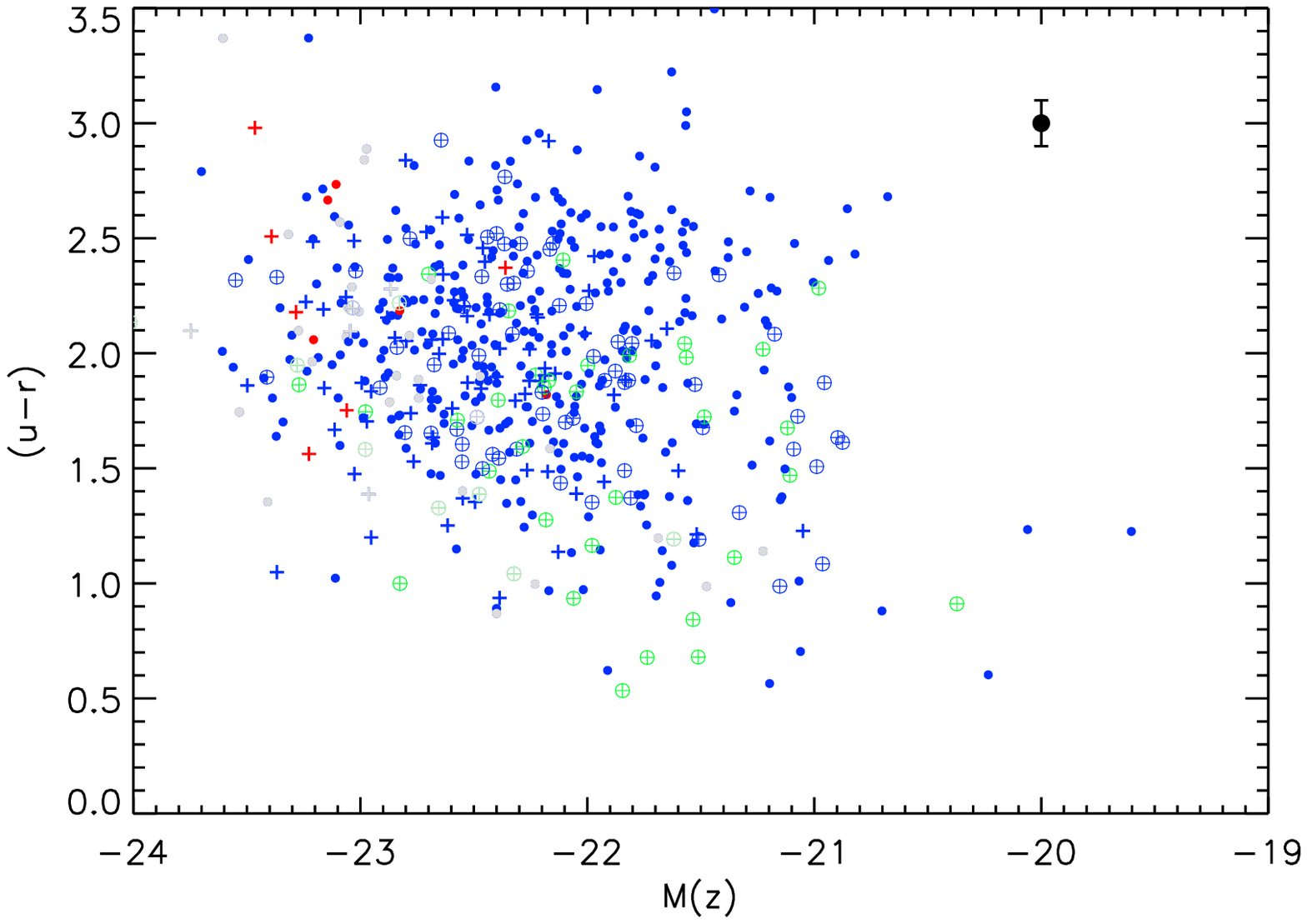}\\
\end{array}$
\caption{TOP: The $(NUV-r)$ colour-magnitude relation (CMR). The
NUV filter is centred at $2300\AA$. BOTTOM: The optical $(u-r)$
CMR. Morphologies are shown colour-coded and symbol types
correlate with the interaction status of the object (see Section
2.2 for details). The dotted line in the top panel indicates the
position of the UV `red sequence', taken from a study of the
nearby early-type population (Kaviraj et al. 2007b). The
morphological classifications were performed only for objects with
$z<0.15$ (see Section 2.2 for details). Galaxies outside this
redshift range are shown in grey.} \label{fig:cmrs}
\end{center}
\end{figure}

Since {\color{black} the model library includes a large range of
dust values} and the UV colours are sensitive to dust, we wish to
avoid including models in our parameter space where a spuriously
large RSF fraction (i.e. $f_2$) may fit the UV colours of a galaxy
due to an unreasonably high dust extinction. {\color{black}This is
not just desirable but also necessary, since the results of
Burgarella et al. (2005) imply that the nearby LIRG sample
satisfies $A(FUV)\sim2.7$ (which translates to $E_{B-V}\sim0.3$
using a standard Calzetti dust law)}. Since the available values
of $E_{B-V}$ in the model library spans $0<E_{B-V}<1$, we apply a
prior to the $E_{B-V}$ parameter space to ensure realistic values
of dust extinction {\color{black} are employed for each LIRG}.

{\color{black}In the local Universe, the obscuration in
star-forming galaxies correlates with the FIR luminosity
\citep[e.g.][]{Calzetti1995,Wang1996,Adelberger2000,Hopkins2001}.
Following \citet[][see their Eqns. 1, 2 and 3]{Hopkins2001} we
estimate the dust extinction in the nebular emission lines
($E^{G}_{B-V}$) for each galaxy from the FIR luminosity and
convert it to a value appropriate for the stellar continuum
$E^{S}_{B-V}$ using $E^{S}_{B-V} = 0.44 \times E^{G}_{B-V}$
\citep{Calzetti2000}. The prior on the dust is applied, for each
galaxy, by restricting the model library to $E^{S}_{B-V}\pm0.1$.
It is worth noting that the applied prior is not particularly
strong. A large range of \emph{realistic} dust values is permitted
for each galaxy \emph{but the E(B-V) parameter space typically
greater $E^{S}_{B-V}\sim0.5$ is discarded}.

Converting the values of $E^{S}_{B-V}$ derived using the procedure
described above to $A_{FUV}$ using a Calzetti law yields
$1.8<A_{FUV}<3.3$ for our sample of LIRGs. This compares very well
with the recent LIRG study of Burgarella et al. (2005a) who find a
median obscuration of $A_{FUV} = 2.7 \pm 0.8$, using theoretical
calibrations derived from GALEX and IRAS data by Buat et al.
(2005). Note that Buat et al.'s calculations are resistant to the
choice of the exact extinction curve (see their Section 3.1).}

The principal free parameters ($t_1$, $t_2$, $f_2$ and $\tau_2$)
are estimated by comparing each observed galaxy to every model in
the synthetic library, with the likelihood of each model ($\exp
-\chi^2/2$) calculated using the value of $\chi^2$ computed in the
standard way. From the joint probability distribution, each
parameter is marginalised to extract its one-dimensional
probability density function (PDF). We take the median of this PDF
as the best estimate of the parameter in question and the 16 and
84 percentile values as the uncertainties on this estimate.

For galaxies which do not have $FUV$ detections, the parameter
estimation is performed using only the $NUV$, $u, g, r, i$ and
$z$-band filters. We do not use the $FUV$ detection limit as an
upper limit to the $FUV$ flux, primarily because the uncertainties
on the estimated parameter effectively become unbounded. Previous
work on early-type galaxies and E+A systems indicates that
addition of the $FUV$ filter does not affect the median values but
does reduce the size of the uncertainties. The cosmological
parameters used in this study assume a $\Lambda$CDM model:
$h=0.7$, $\Omega_{m}=0.3$ and $\Omega_{\Lambda}=0.7$.


\section{Star formation histories}
It is instructive to look first at the UV and optical
colour-magnitude relations of the LIRG population, since the
derived SFHs are effectively driven by the observed colours
(Figure \ref{fig:cmrs}). Morphologies are shown colour-coded and
symbol types correlate with the interaction status of the object.
The dotted line in the top panel indicates the position of the UV
`red sequence', taken from a study of the nearby early-type
population (Kaviraj et al. 2007b). Not unexpectedly, virtually all
LIRGs lie away from the red sequence, notwithstanding their
relatively high dust content. The bulk of the sample occupies the
`blue cloud' ($NUV-r<3.5$). We note that the distributions in both
the UV and optical colours are very broad and that smaller
galaxies are not preferentially bluer as typically indicated by
the downsizing phenomenon \citep[see e.g.][]{Cowie1996}. However,
the star formation regimes in these objects are significantly more
severe than in the normal galaxy population and the observed
colours are dominated by the vigorous RSF being experienced by
these systems. Since the downsizing phenomenon reflects the
make-up of the underlying stellar population of galaxies that
forms over the lifetime of the Universe, it is reasonable to
expect these trends to be temporarily washed out by the strong RSF
in the LIRG sample.

\begin{figure*}
\begin{minipage}{172mm}
\begin{center}
$\begin{array}{cc}
\includegraphics[width=3.5in]{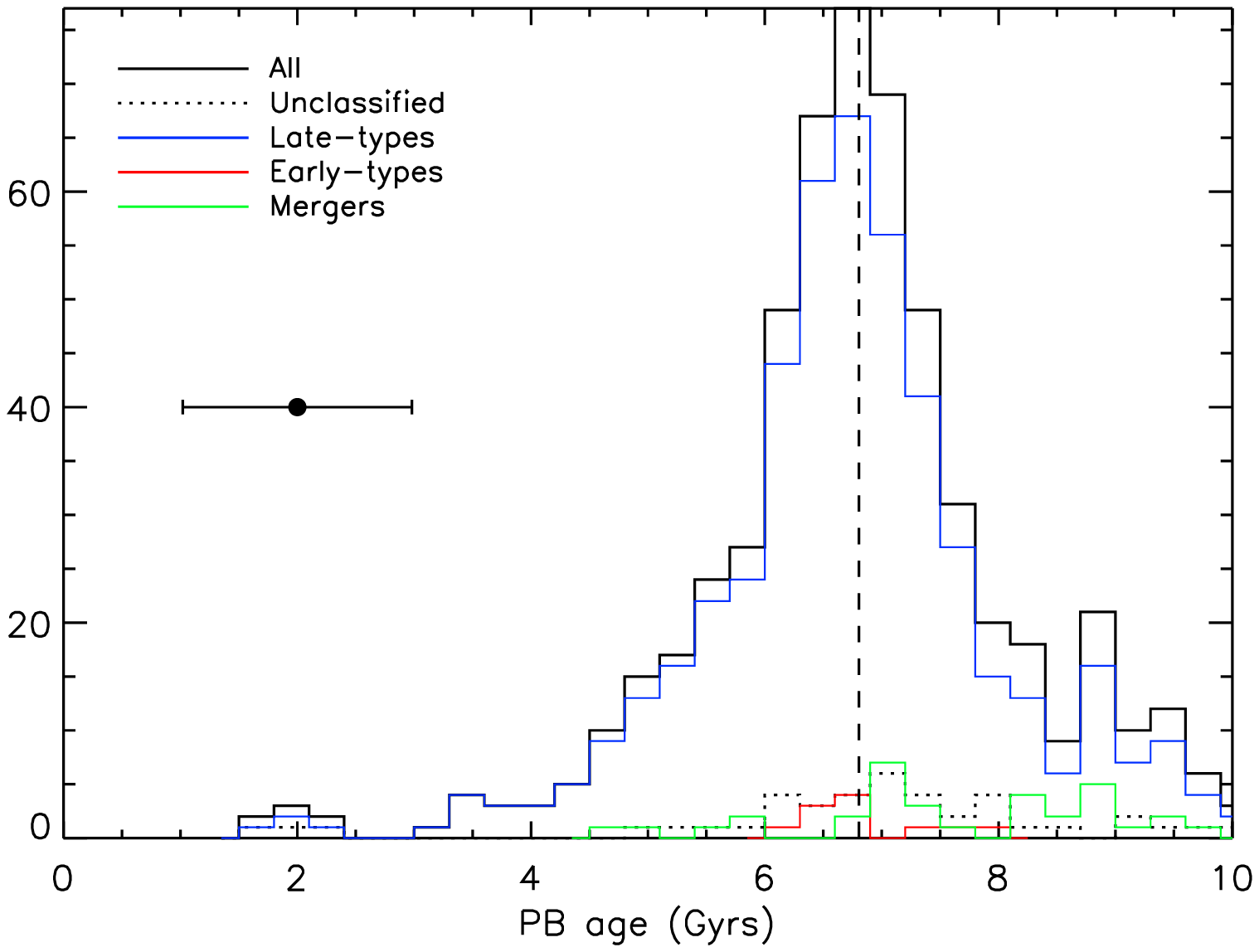}&\includegraphics[width=3.5in]{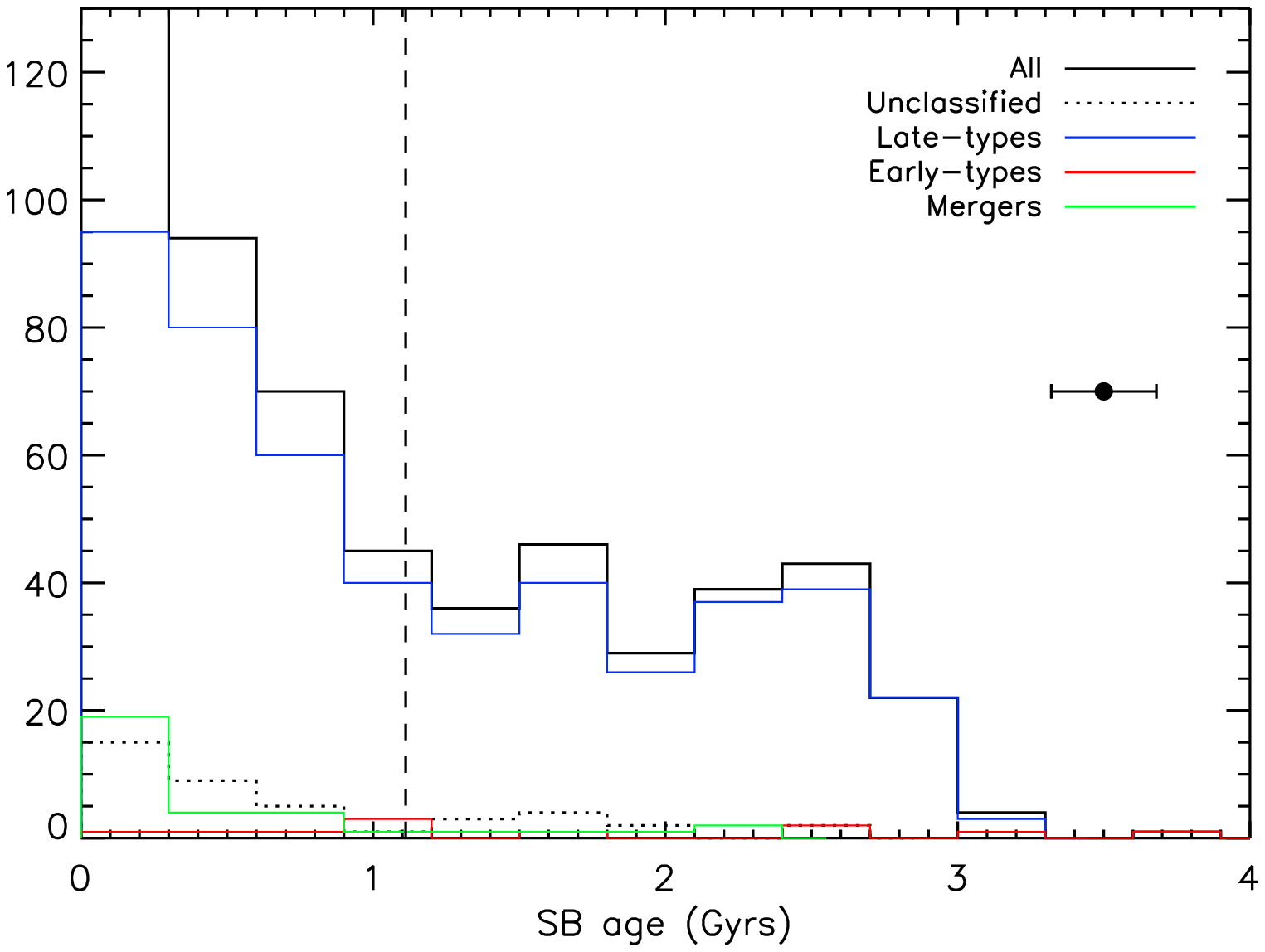}\\
\includegraphics[width=3.5in]{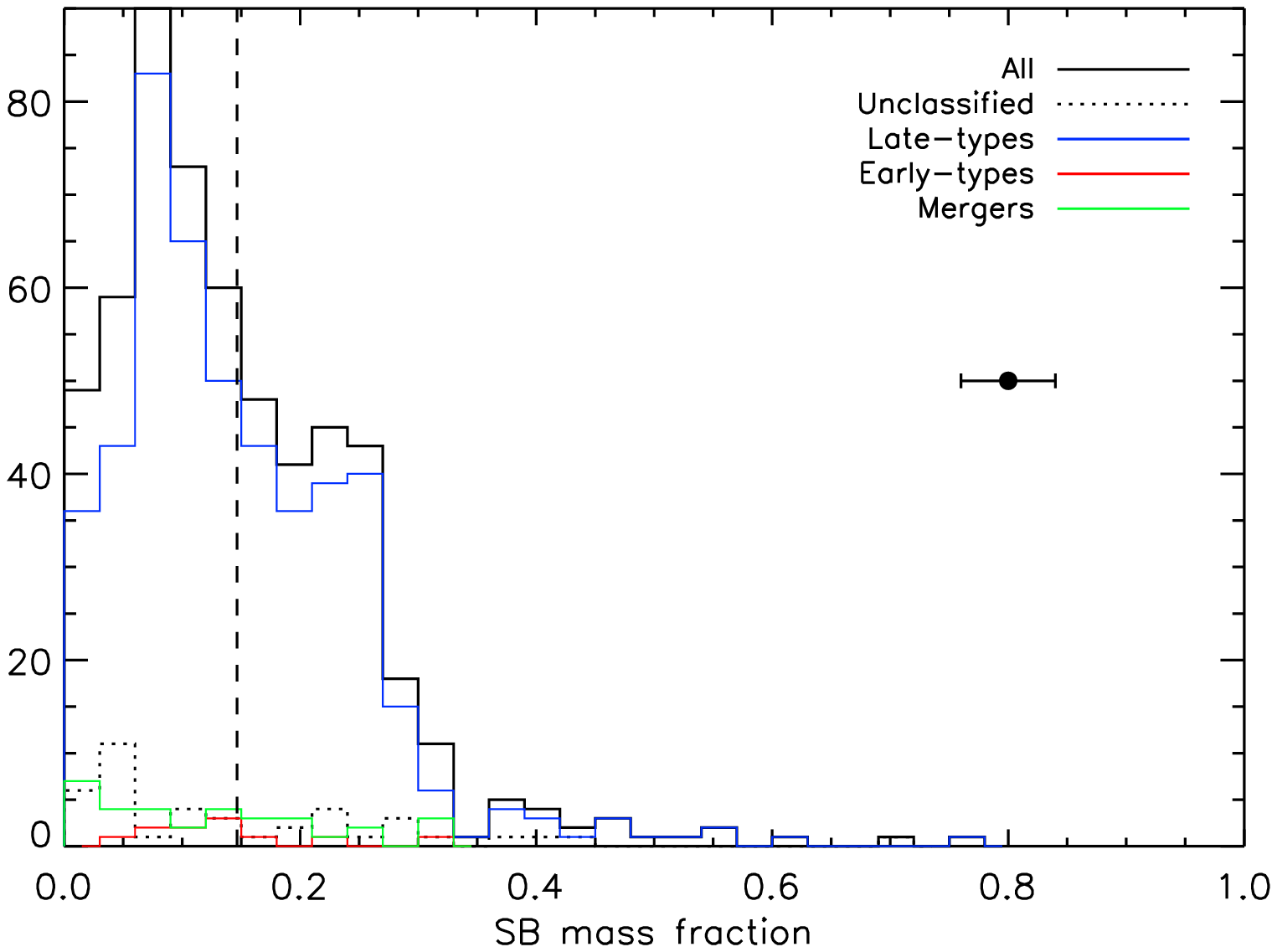}&\includegraphics[width=3.5in]{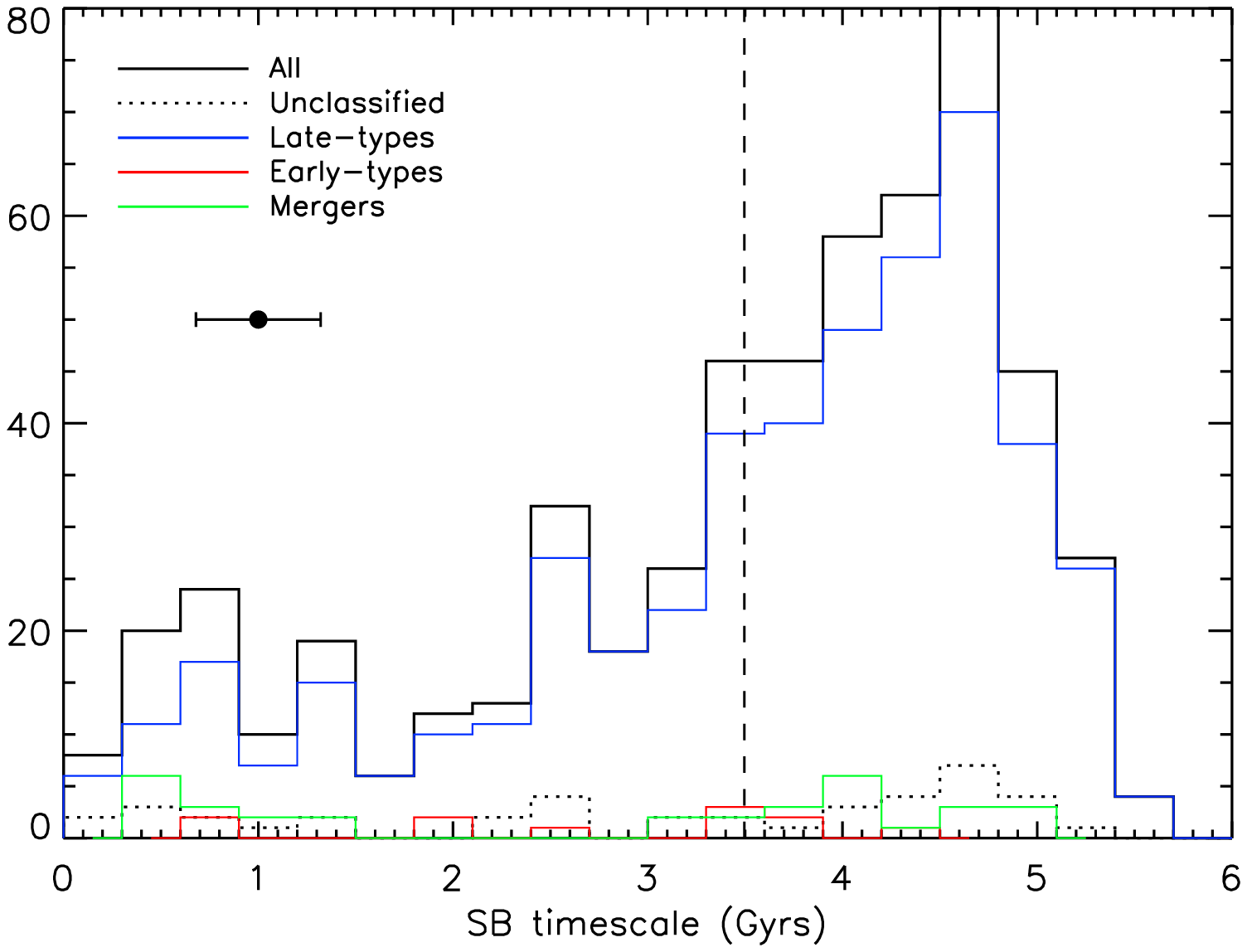}
\end{array}$
\caption{The derived SFHs of the LIRG sample as a function of the
morphology of the galaxies. TOP-LEFT: The primary burst ages
($t_1$). TOP-RIGHT: The secondary burst ages ($t_2$). BOTTOM-LEFT:
The secondary burst mass fractions ($f_2$). BOTTOM-RIGHT: The
secondary burst timescales ($\tau_2$). The mean value of each
parameter for the population as whole is indicated by the vertical
dashed lines. The morphological classifications were performed
only for objects with $z<0.15$ (see Section 2.2 for details). The
`unclassified' category contains all objects outside this redshift
range.} \label{fig:sfh_morphology}
\end{center}
\end{minipage}
\end{figure*}

We now present the derived SFHs for galaxies in the LIRG sample.
We explore these SFHs, both as a function of the morphology of the
galaxies, as well as their interaction status. In Figure
\ref{fig:sfh_morphology} we begin by presenting the derived SFH
parameters, split by the morphology of the galaxies. The top-left,
top-right, bottom-left and bottom-right panels show the
distributions of PB ages ($t_1$), SB ages ($t_2$), SB mass
fractions ($f_2$) and SB timescales ($\tau_2$) respectively.

The distribution of PB ages ($t_1$) peaks around $\sim7$ Gyrs,
with the bulk of the sample satisfying $5<t_1<9$ Gyrs (the average
value is $\sim6.8$ Gyrs). Recall that the values of $t_1$ reflect
the (SSP-weighted) \emph{average} ages of the underlying stellar
populations. Note that, since the UV is sensitive only to
populations with ages $<2$ Gyrs, the estimates of $t_1$ are driven
primarily by the optical filters. Given their lower sensitivity to
young stellar populations, the uncertainties in the values of
$t_1$ are high ($\sim1.2$ Gyrs). Nevertheless, it is worth noting
that very few of the galaxies have underlying populations that are
well-fit by purely old ($>10$ Gyrs) stellar populations. This
result is not unexpected, since the LIRG sample is dominated by
late-type galaxies, which have typically formed stars over the
lifetime of the Universe.

The top-right panel of Figure \ref{fig:sfh_morphology} indicates
that roughly 60\% of the LIRGs in this study began their RSF
episode within the last Gyr (i.e. SB age or $t_2<1$ Gyr), while
for the remaining galaxies $1<t_2<3$ Gyrs. Virtually none of the
objects exhibit $t_2\gtrsim3$. The mergers (green) have a
distribution in $t_2$ that is sharply peaked below $t_2<0.5$ Gyrs,
consistent with the fact that their RSF episodes have begun
recently or, at least, are still in progress. The bottom-left
panel of Figure \ref{fig:sfh_morphology} indicates that nearby
LIRGs have formed up to 35\% of their stellar mass in the recent
star formation episode. The mean value is $\sim15\%$. Not
unexpectedly, the mergers (green) peak towards lower mass
fractions, since the SF episode is less advanced than in the other
morphological types. The bottom-right panel of Figure
\ref{fig:sfh_morphology} indicates that the SB timescales
($\tau_2$) involved in these events are typically large, of the
order of a few Gyrs, indicating that the SFR does not decline
significantly during the course of the RSF episode.



\begin{figure}
\begin{center}
\includegraphics[width=3.5in]{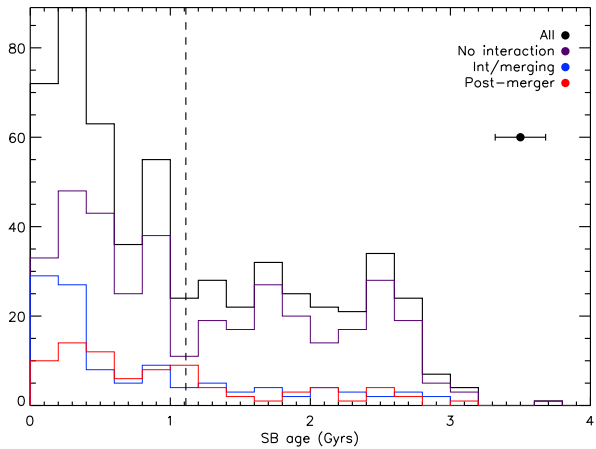}
\caption{The secondary burst ages ($t_2$) as a function of the
interaction status of the LIRGs in our sample.}
\label{fig:sfh_interactionstatus}
\end{center}
\end{figure}

In Figure \ref{fig:sfh_interactionstatus}, we briefly present the
SB ages of the LIRG sample as a function of their interaction
status. The PB ages ($t_1$), SB mass fractions ($f_2$) and SB
timescales ($\tau_2$) are omitted for clarity, as all categories
(isolated/interacting/post-mergers) have similar distributions in
these variables, akin to the ones presented in Figure
\ref{fig:sfh_morphology}. We find that around 74\% of the galaxies
that are currently interacting systems began their RSF episodes
within the last 0.5 Gyrs. Recall that the `interacting/merging'
category includes all systems that are currently interacting, not
just mergers which are a special case where two systems are
partially fused.

\begin{figure}
\begin{center}
\includegraphics[width=3.5in]{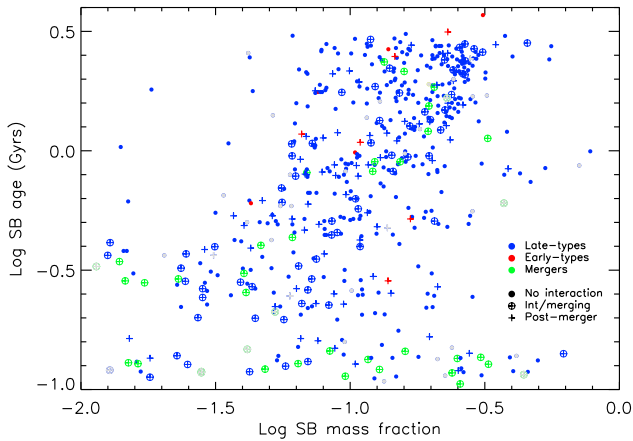}
\caption{The secondary burst ages ($t_2$) and mass fractions
($f_2$) of our LIRG sample. The morphology of the galaxies is
shown colour-coded and their interaction status is shown using
different symbol types. Note that the morphological
classifications were performed only for objects with $z<0.15$ (see
Section 2.2 for details). Objects outside this redshift range,
which were not part of the morphological classification, are shown
in grey.} \label{fig:sbage_mf}
\end{center}
\end{figure}

Finally, in Figure \ref{fig:sbage_mf}, we summarise the SB ages
and SB mass fractions of individual galaxies in our sample.
Morphologies are shown colour-coded and symbol types indicate the
interaction status of each galaxy. Since the morphological
classifications were performed only for objects with $z<0.15$ (see
Section 2.2), objects outside this redshift range are shown in
grey.

\section{The role of AGN}
Several lines of theoretical and observational evidence indicates
that AGN, driven by supermassive black holes that inhabit galactic
centres, are likely to play a significant role in the formation
and evolution of their host galaxies. The principal impact of AGN
is energetic feedback on the inter-stellar medium (ISM) which acts
to remove (cold) gas from the potential well halting star
formation in massive galaxies
\citep[e.g.][]{Silk1998,Benson2003,Silk2005}. Such negative
feedback from AGN is commonly invoked in galaxy formation models
to reproduce the cut-off in the galaxy luminosity function at high
luminosities \citep[][]{Benson2003} and to fit the observed
colours of massive galaxies at present-day
\citep{Kaviraj2005a,deLucia2006}. Without such feedback, the
potentially plentiful supply of cold gas in massive galaxies
results in objects that are predicted to be too massive and too
blue to fit the observations in the nearby Universe. Since
low-redshift LIRGs represent some of the most vigorously star
forming galaxies in the nearby Universe, it is instructive to
explore the potential role of AGN in these systems and search for
tell-tale signs of feedback in their SFHs.

\subsection{Number fractions and morphologies of AGN hosts}
We begin by exploring the proportion of AGN in the LIRG sample
studied in this paper. The overall fraction of AGN is $\sim14\%$.
We find that the AGN fraction shows a dependence on the
interaction status of the galaxies - it rises from 13.1\% in
non-interacting objects to 17.6\% in the interacting population
and back down to 13.4\% in the post-mergers, strongly consistent
with the idea that interactions trigger the onset of AGN. A
similar result has been found recently by \citet{Alonso2007} in
nearby close pairs drawn from the SDSS. Their results indicate
that the AGN fraction in close pairs that exhibit definite signs
of interactions is $\sim4\%$ higher than in pairs that are not
interacting. This value is strikingly similar to the corresponding
statistics found in our LIRG sample. {\color{black}We should note,
however, that the number of galaxies in the non-interacting and
post-merger categories is fairly small (of the order of a 100
objects each). Thus, while the trend in the AGN fraction is
qualitatively similar to previous works, the low number statistics
make it difficult to ascertain how significant the change in the
AGN fraction really is.}

\begin{figure}
\begin{center}
\includegraphics[width=3.5in]{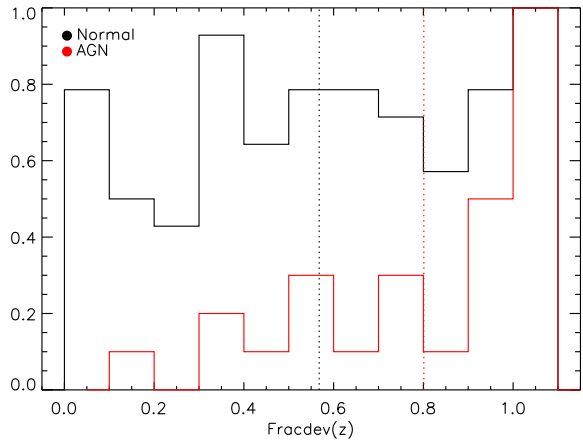}
\caption{Comparison of the bulge/disk morphologies of LIRGs that
host AGN (red), to those that show composite spectra (green) and
those that do not show any indication of an AGN (black). Note that
we restrict the sample to nearby ($0.06<z<0.09$) objects for this
comparison, to ensure the accuracy of the light profiles and
explore the \texttt{fracdev} parameter in the $z$-band filter
which is least affected by the RSF and therefore traces the bulk
stellar population of the galaxy most faithfully.}
\label{fig:fracdevs}
\end{center}
\end{figure}

Previous studies of the AGN population have indicated that they
typically reside in massive bulge-dominated systems (Kauffmann et
al. 2003a). While our sample of LIRGs is a subset of the galaxy
populations studied before, it is instructive to look at the
distribution of bulge/disk morphologies of AGN in these galaxies,
since they are far more extreme in terms of their star formation
properties than the normal galaxy population. We compare the
bulge/disk morphologies of the AGN hosts in this sample to
galaxies without AGN using the SDSS \texttt{fracdev} parameter.
The SDSS pipeline fits both a deVaucouleur's and an exponential
model to the light profile of each galaxy in the $u,g,r,i,z$
filters. A composite \emph{best-fit} is then generated using a
linear combination of the two fits. \texttt{fracdev} is the weight
of the deVaucouleur's model in this composite fit and provides a
quantitative measure of the bulge component of the galaxy.
Early-type galaxies in the nearby Universe typically have
\texttt{fracdev}$>0.85$.

In Figure \ref{fig:fracdevs} we compare the \texttt{fracdev}
distribution of the AGN hosts to that in the normal galaxies. Note
that we restrict the sample to nearby ($0.06<z<0.09$) objects to
ensure the accuracy of the fitted light profiles and explore the
\texttt{fracdev} parameter in the $z$-band filter which is least
affected by the RSF and therefore traces the bulk stellar
population of the galaxy most faithfully. We find that, consistent
with previous results, the distribution of the AGN hosts is
strongly skewed towards higher values of \texttt{fracdev}, while
the normal galaxy population shows a flat distribution in this
parameter. We conclude that AGN, even in the most strongly
star-forming systems in the nearby Universe, occupy galaxies which
have a dominant bulge component.


\subsection{Colours and star formation histories of the AGN hosts}
Given their significance in galaxy formation models, it is useful
to constrain the role of AGN in the LIRG population studied here.
We wish to explore (a) the time after the onset of star formation
when the AGN typically appears and (b) whether there is any
evidence of negative feedback from AGN in their host galaxies.
Both these aims can be achieved by comparing the derived SFHs of
the AGN hosts to that of the normal galaxies.

\begin{figure}
\begin{center}
$\begin{array}{c}
\includegraphics[width=3.5in]{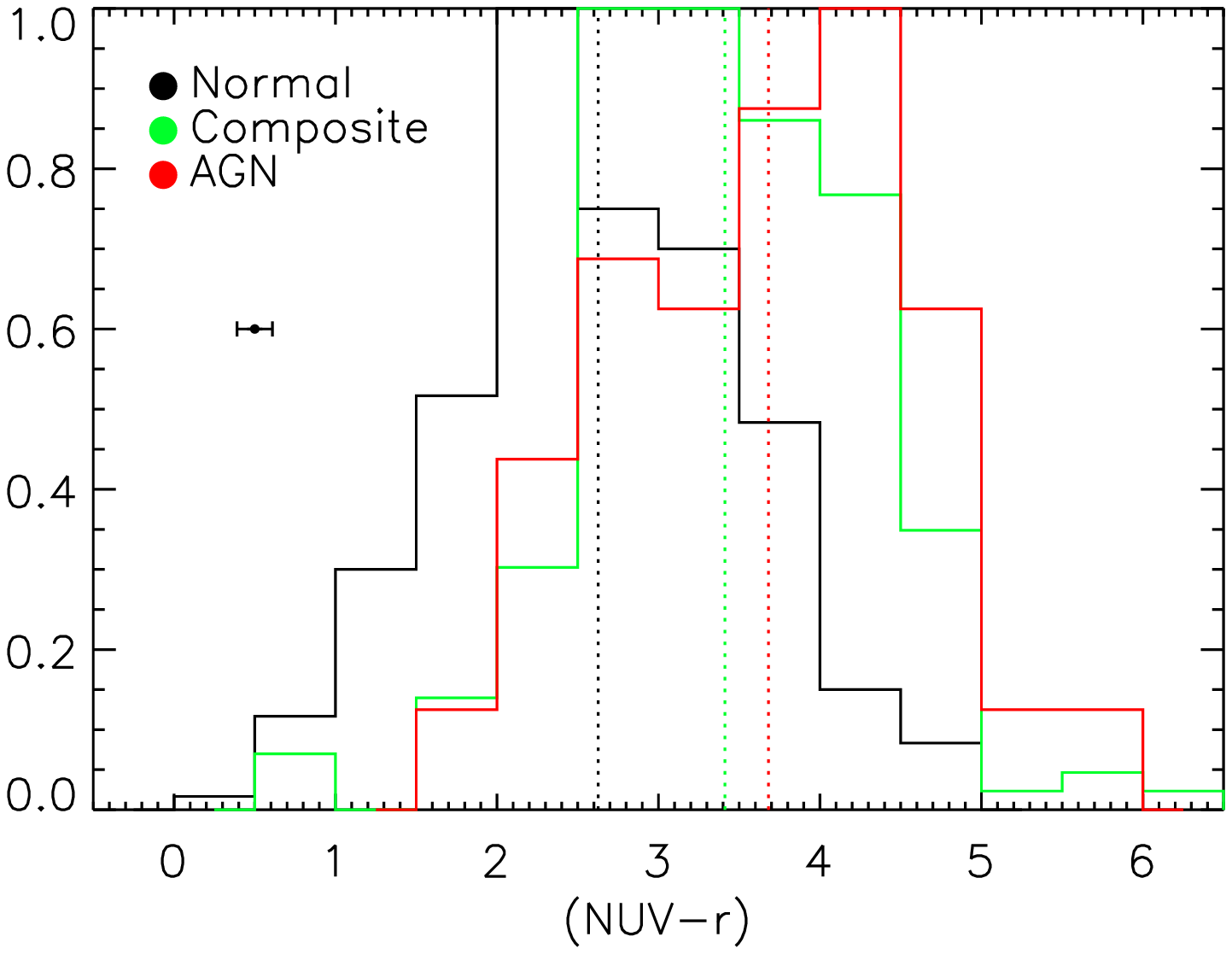}\\
\includegraphics[width=3.5in]{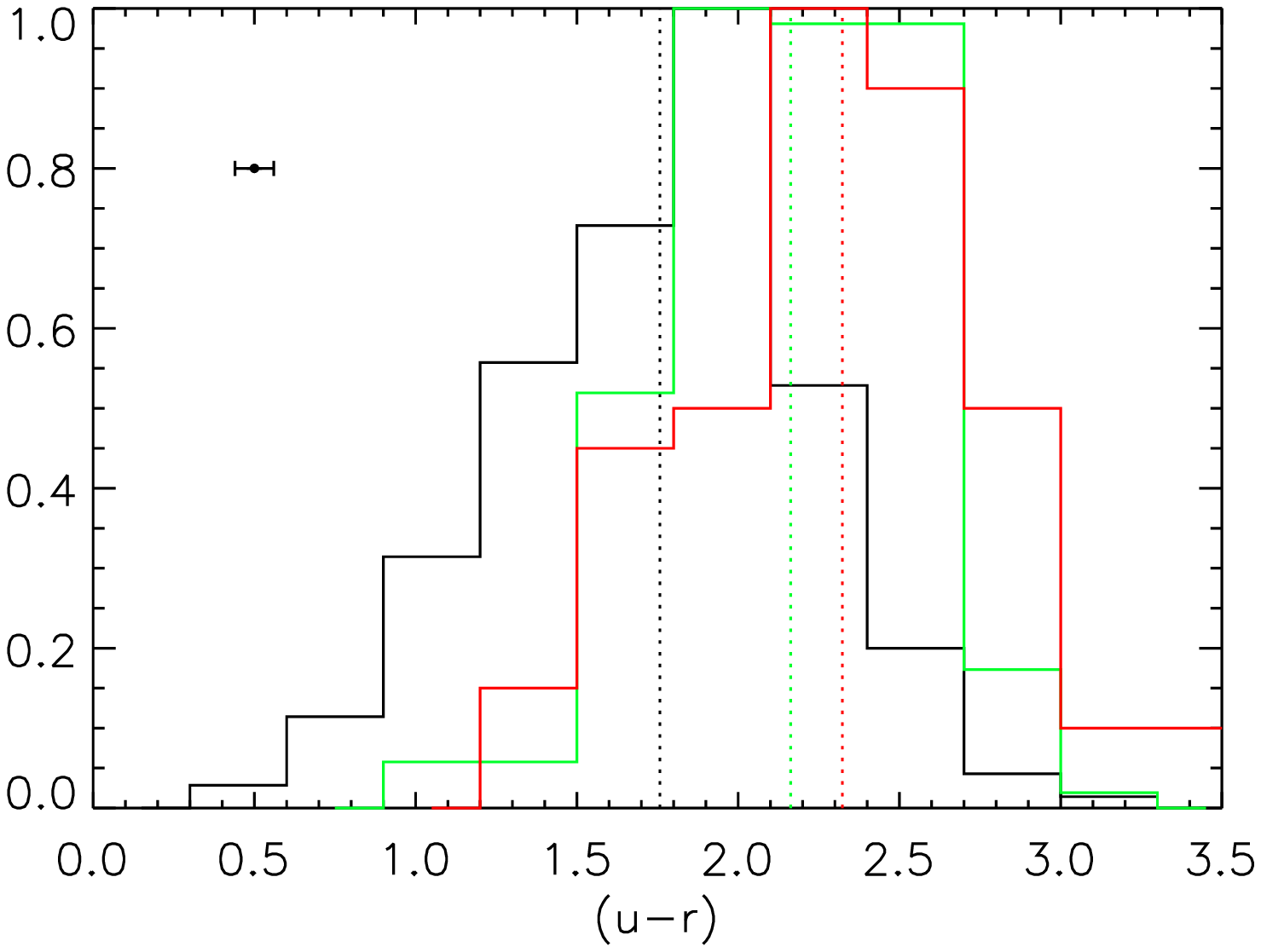}\\
\includegraphics[width=3.5in]{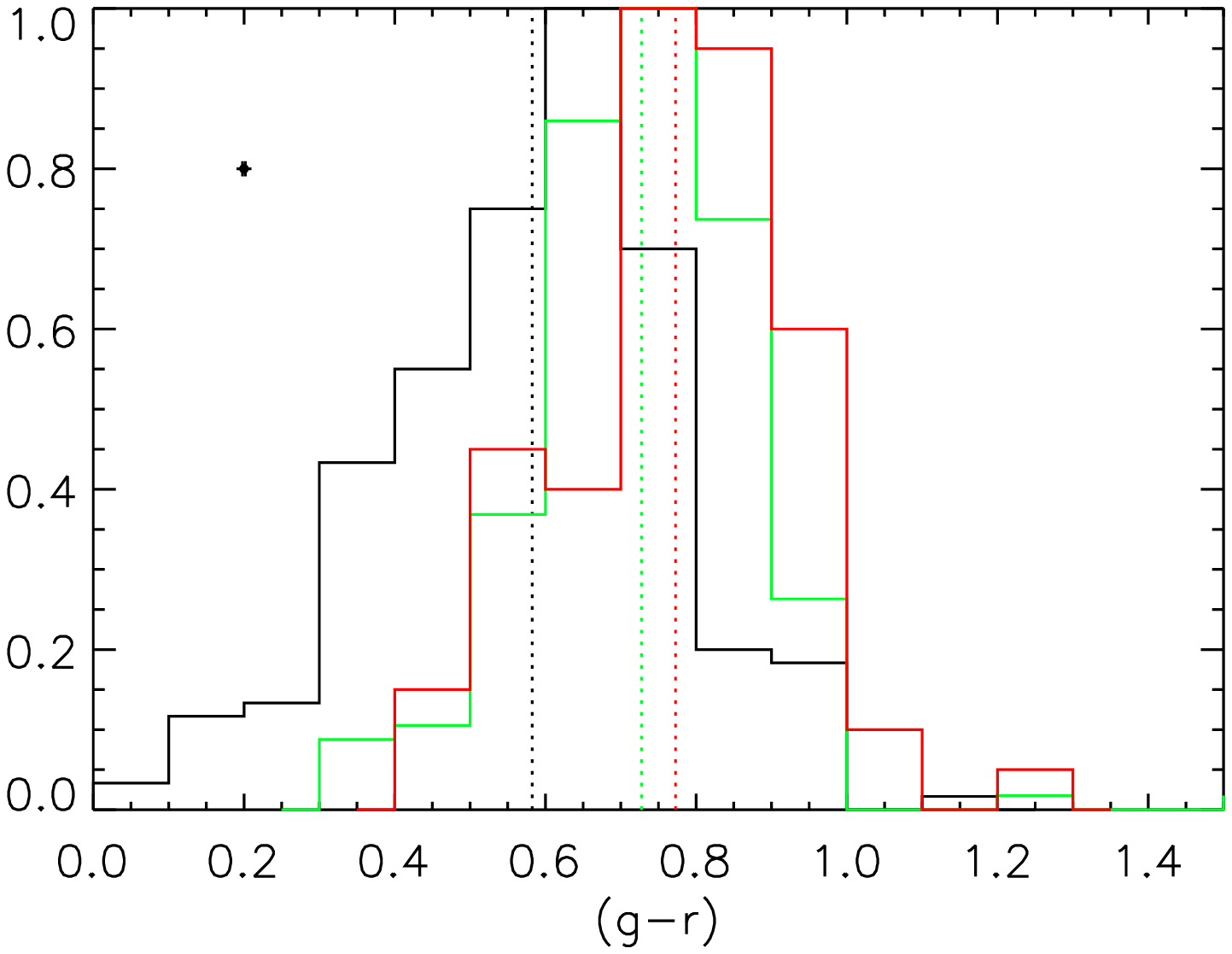}
\end{array}$
\caption{Comparison of the colours of LIRGs that host AGN (red),
to those that show composite spectra (green) and those that do not
show any indication of an AGN (black).}
\label{fig:agn_colours}
\end{center}
\end{figure}

In Figure \ref{fig:agn_colours} we compare the colours of the AGN
hosts (red), to those that show composite spectra (green) and
galaxies that do not show any indication of an AGN (black). We
find that, irrespective of the colour, the AGN hosts are
consistently redder than the normal galaxies. The mean values
(indicated using dotted lines) show that AGN are redder in
$(NUV-r)$, $(u-r)$ and $(g-r)$ by $\sim1.2$, $\sim0.6$ and
$\sim0.4$ mags respectively. The systematic reddening persists if
we split the sample into luminosity bins and/or explore different
redshift ranges. The reddening in colour could be produced either
by an enhanced dust content in AGN or by a systematic difference
in their SFHs. Clearly both explanations have to be explored
before the colour discrepancy can be explained.

\begin{figure}
\begin{center}
$\begin{array}{c}
\includegraphics[width=3.5in]{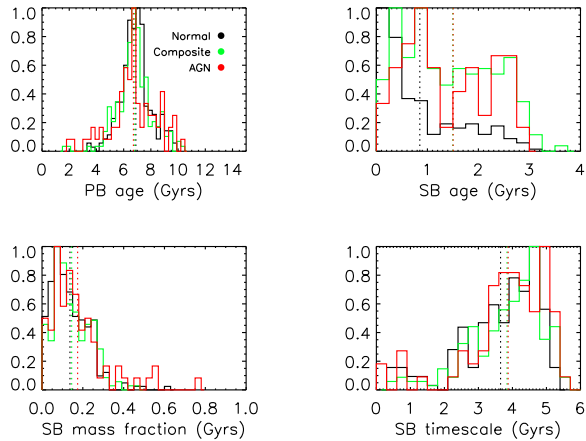}
\end{array}$
\caption{Comparison of the SFHs of AGN, composites and normal
galaxies in the LIRG sample. Note that the histograms for all
populations have been normalised to 1. The fraction of AGN in the
LIRG sample studied here is $\sim14\%$.}
\label{fig:sfh_differential}
\end{center}
\end{figure}

We find that the distributions of $E_{B-V}$ do not indicate a
systematic dust enhancement in the AGN hosts
\footnote{{\color{black}In fact, the discrepancies in the colours
themselves do not favour a differential dust content as an
explanation for the redder colours in the AGN hosts. While a small
differential dust extinction can, in principle, produce a large
offset in the UV colour (because e.g. $A_{(NUV-r)} \sim 6 \times
E_{B-V}$), the extinction in the $(g-r)$ colour is $A_{(g-r)} \sim
1.09 \times E_{B-V}$ using the Calzetti dust law. If the SFHs of
the AGN hosts and the normal galaxies were identical, it is
difficult to explain the observed average discrepancy in $(g-r)$
($\sim0.4$ mags), by simply invoking a systematic $E_{B-V}$ offset
of $\sim0.4$ mags.}} leading us to consider the alternative
explanation for the reddening in the AGN colours - that they are
driven by a systematic difference in their SFHs. In Figure
\ref{fig:sfh_differential} we compare the derived SFH parameters
of the AGN, composites and normal galaxies. We find that the
distributions of PB ages (top-left panel) are virtually identical
for all categories, implying that the underlying populations of
the galaxies are very similar. The SB timescales are also
consistent between the three categories (bottom-right panel). The
principal difference in the SFHs is that the SB ages of the AGN
hosts show an offset of $\sim0.6$ Gyrs compared to the normal
galaxies. In particular, the distribution of SB ages in the AGN
declines dramatically at values less than $\sim0.5-0.7$ Gyrs.

These results indicate that the SFHs in AGN do not differ
significantly from those in the normal galaxies but that we are
simply observing the AGN hosts a \emph{longer} time after the
onset of star formation (since the SB ages are larger). The rapid
decline in the number of AGN at SB ages less than $\sim0.5-0.7$
Gyrs suggests that AGN typically appear $\sim0.5-0.7$ Gyrs after
the onset of star formation. It is worth noting that a similar
time delay (a few hundred Myrs) between the peak of the star
formation and the rise of the AGN has been reported in a recent
study of AGN activity in early-type galaxies
\citep{Schawinski2007}.

The similarity between the properties of the star formation in the
AGN hosts and the normal galaxies indicates that the AGN itself
does not have a measurable impact on the SFHs of their host
galaxies. Negative feedback can be expected to remove or heat cold
gas from the potential well, leading to a reduction in the star
formation activity. If star formation is strongly quenched, we
would expect the decay timescale of the star formation activity to
decrease, as the SFR is strongly reduced by removal of cold gas
from the system. This, in turn, will be reflected in a reduction
of the SB timescales. Such an effect on the SB timescales is
clearly seen in post-starburst (E+A) galaxies, where the lack of
H$\alpha$ in emission, coupled with deep Balmer lines, indicates
that a vigorous period of star formation has been rapidly quenched
(Kaviraj et al. 2007c). However, as Figure
\ref{fig:sfh_differential} indicates, we find no evidence for a
similar effect in the LIRG population. In fact the average SB
timescales (see the vertical dotted lines) in the AGN hosts are
marginally larger.

\begin{figure*}
\begin{minipage}{172mm}
\begin{center}
$\begin{array}{ccc}
\includegraphics[width=2.3in]{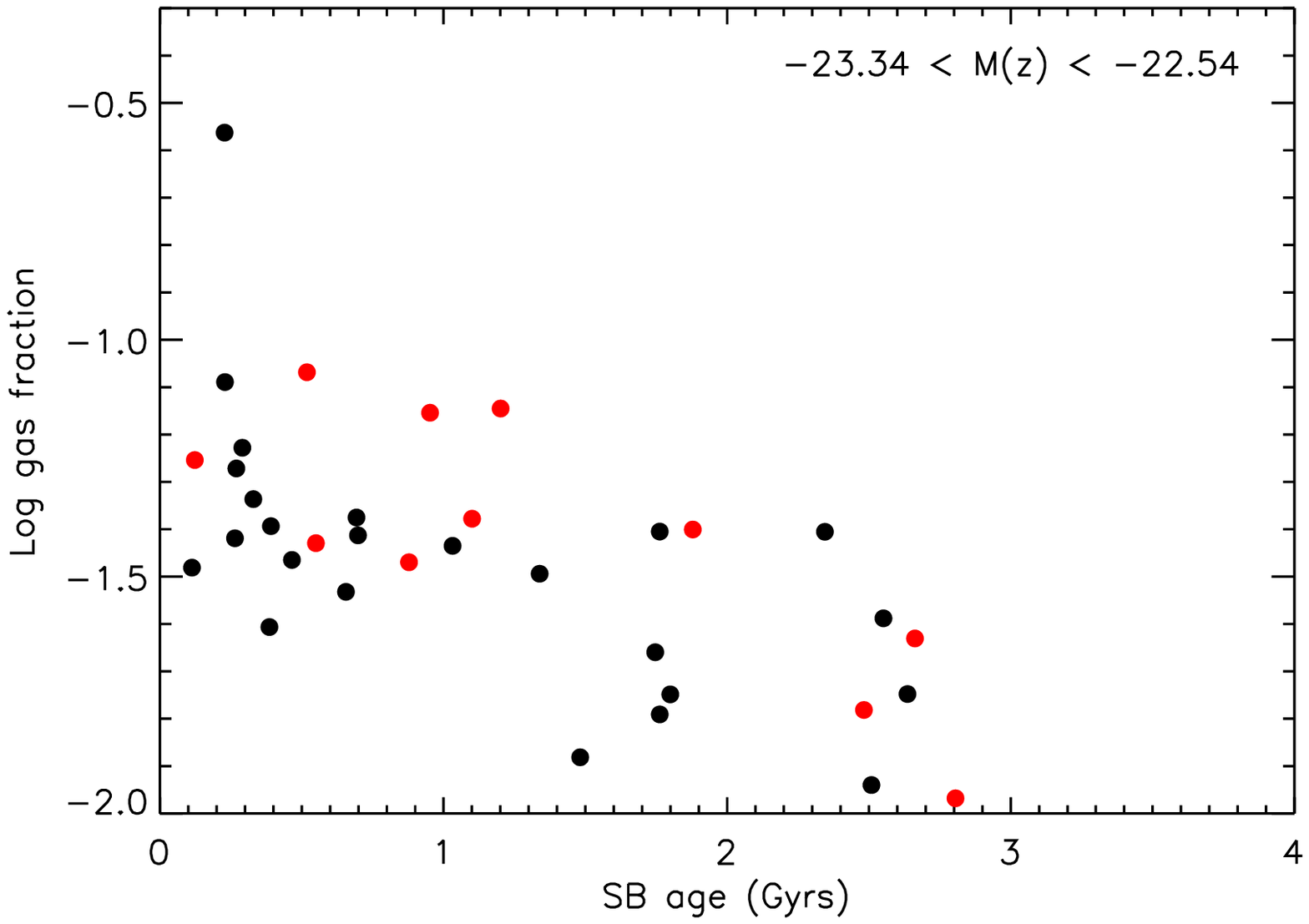}&\includegraphics[width=2.3in]{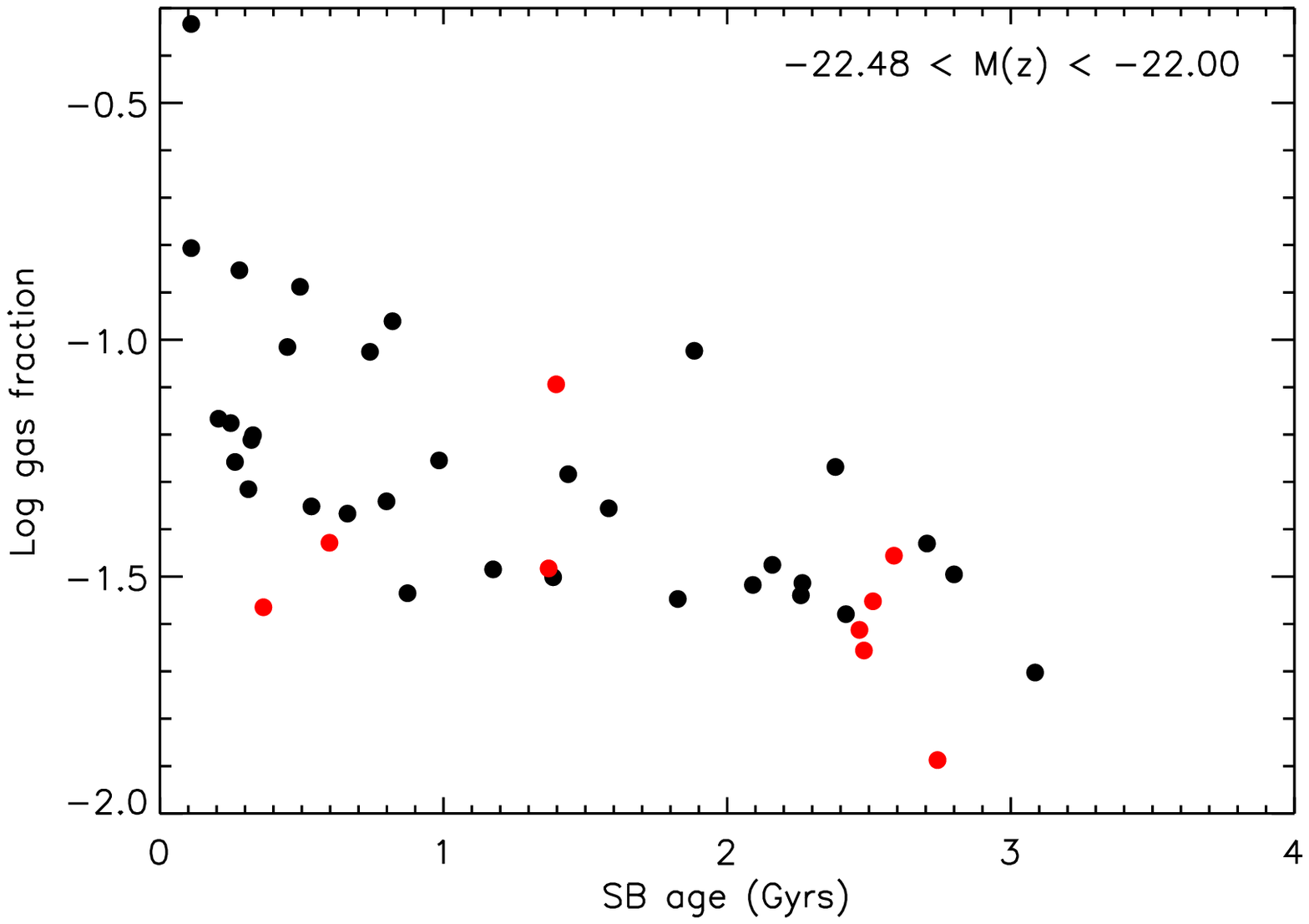}&\includegraphics[width=2.3in]{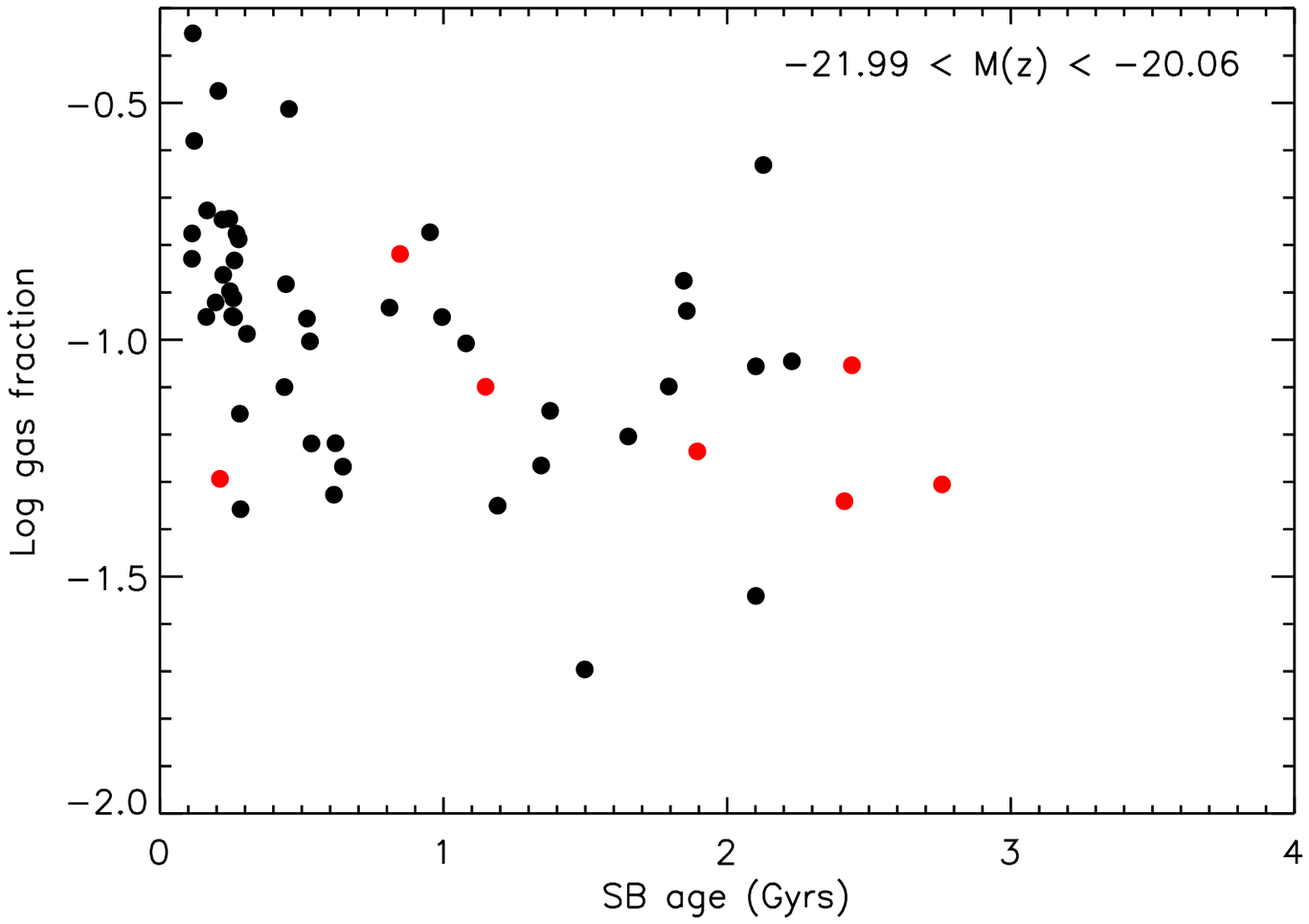}
\end{array}$
\caption{The evolution of the cold gas fraction as a function of
the age of the secondary burst (i.e. the SB age parameter) in the
AGN hosts (red) and normal galaxies. Note that, since the cold gas
fraction decreases with the luminosity of the galaxy \citep[see
Figure 5 in][]{Wang2006}, we bin the sample into three luminosity
bins in $M(z)$ with roughly equal numbers of objects and consider
a narrow redshift range ($0.06<z<0.09$).} \label{fig:gasfractions}
\end{center}
\end{minipage}
\end{figure*}

As a further check of this result, we compare the evolution of the
cold gas fractions in the AGN hosts to that in the normal galaxies
(Figure \ref{fig:gasfractions}). Following \citet{Wang2006} we
estimate the cold gas mass using the correlation between the CO
and IR luminosities derived by Gao \& Solomon (2004a, by combining
their Eqns 1 and 2), and apply the `standard'
CO-to-H$_{\textnormal{2}}$ ratio estimated from giant
molecular clouds in the disk of the Milky Way (see Eqn. 5 in Gao
\& Solomon 2004b). Note that, since there is an evolution in the
gas fraction as a function of mass/luminosity in the LIRG
population \citep[see Figure 5 in][]{Wang2006}, we split Figure
\ref{fig:gasfractions} into three luminosity bins in $M(z)$ with
roughly equal numbers of objects.

If the AGN are depleting the cold gas reservoirs in their host
galaxies then we would expect to see - \emph{at a given value of
SB age} - a systematically lower cold gas content in the AGN hosts
than in the normal galaxies. Figure \ref{fig:gasfractions}
indicates that, as should be expected, the cold gas fractions show
a steady decline with SB age, regardless of the luminosity bin
being considered. However, it is also apparent that the AGN do not
show a preferentially lower cold gas content and that the trends
in the AGN are indistinguishable from those in the normal
galaxies. The lack of discrepancy in the cold gas content
\emph{independently} supports the observed similarity in the SFHs
of the AGN and their normal counterparts. We conclude, therefore,
that there is no measurable evidence for feedback in LIRGs that
host Type 2 AGN, either in terms of cold gas removal or the impact
on the SFH that would result from such gas depletion.

\section{LIRGs with spheroidal morphologies - mixed major mergers?}
The sample of spheroidal objects identified in this study (Figure
\ref{fig:etg_gallery}) allows us to explore the star formation
that is plausibly induced by (gas-rich) major mergers in the
nearby Universe. In this section we summarise the characteristics
of such major merger driven star formation and speculate on the
properties of the progenitors involved, using the derived SFH
parameters of these spheroidal galaxies.

We note first that the early-type galaxies identified in the
visual inspection (shown in red in Figure
\ref{fig:sfh_morphology}) have PB ages ($t_1$) of $\sim6.9$ Gyrs.
Based on the fact that (a) these objects have spheroidal
morphology (b) exhibit LIRG-like SFRs and (c) contain underlying
populations with relatively young average ages, we speculate that
they are the products of `mixed' major mergers where \emph{at
least} one of the progenitors has late-type morphology. This is
because mergers between two ellipticals should result in higher
values of $t_1$, since the red optical colours of the elliptical
population would require both galaxies to have stellar populations
that are overwhelmingly old ($>10$ Gyrs old). Furthermore, since
ellipticals are typically gas-poor systems, a dry major merger (or
equally a minor merger, see e.g. Figure 1 in Kaviraj et al. 2007d)
would not supply enough gas to create the high LIRG-like SFRs in
these systems.

\begin{figure}
\begin{center}
\includegraphics[width=3.5in]{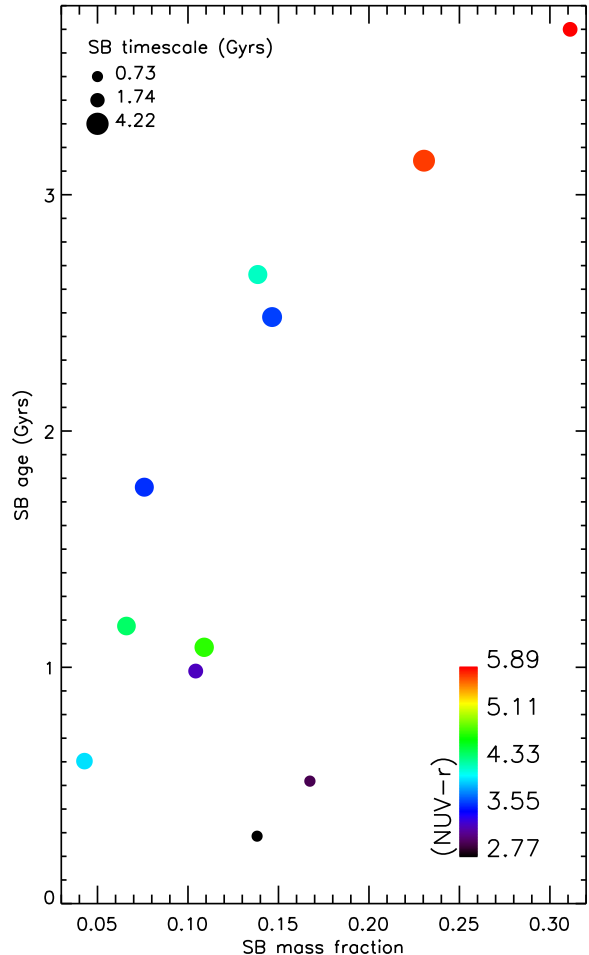}
\caption{The derived SFH parameters of spheroidal objects in the
LIRG sample. See Figure \ref{fig:etg_gallery} for a description of
the sample.} \label{fig:major_mergers}
\end{center}
\end{figure}

In Figure \ref{fig:major_mergers} we summarise the SFHs of the
early-type objects in this sample. The points are colour-coded
using the $(NUV-r)$ colour, symbol sizes correlate with the SB
timescales and galaxy images are placed next to the data points to
demonstrate their morphologies. We find that such `mixed' major
mergers, that produce LIRG-like objects in the local Universe,
typically form between 5 and 30\% of the stellar mass in their
remnants. The duration of these events is between 0.3 and 4 Gyrs,
although, since most of the objects are unrelaxed, these values
are typically lower limits. The star formation timescales range
between 0.7 and 4.2 Gyrs and comparison of the SB ages and SB
timescales in Figure \ref{fig:major_mergers} indicates that the
SFR does not decline significantly through the event.

\section{Summary and conclusions}
We have performed a quantitative study of the SFHs of 561 LIRGs in
the low-redshift ($z<0.2$) Universe. The sample consists of
galaxies in the SDSS DR2 that satisfy the LIRG criterion
($L_{IR}>10^{11}L_{\odot}$) in their IRAS fluxes and are detected
by the GALEX UV space telescope.

A visual inspection of a subsample of galaxies with $r<16.8$ and
$z<0.1$ (for which eyeball classification of galaxy morphologies
is reliable) indicates that the fraction of spheroidal or
near-spheroidal objects that could be progenitors of elliptical
galaxies, is small (4\%). The remaining objects are
morphologically late-type or ongoing mergers. 61\% of the
population do not show any indication of interactions at the
fiducial depth of the SDSS images, while the remaining objects are
either in the process of interacting ($18\%$) or show post-merger
morphologies ($19\%$).

Notwithstanding their high dust content ($<E_{B-V}> \sim 0.3$),
the overwhelming majority of low-redshift LIRGs lie in the UV blue
cloud. We have parametrised the SFH of each galaxy in our sample
by comparing their GALEX (UV) and SDSS (optical) photometry to
synthetic photometry from a large library of $\sim 1.8$ million
model SFHs. We find that the (SSP-weighted) \emph{average} age of
the underlying stellar populations in these objects is typically
between 5 and 9 Gyrs, with a mean value of $\sim6.8$ Gyrs. Roughly
60\% the LIRG population began their recent star formation (RSF)
episode within the last Gyr, while the remaining objects are
consistent with values between 1 and 3 Gyrs in the past. The age
of this recent burst does not exceed $\sim3$ Gyrs in virtually any
of the objects in this sample. LIRGs in the low redshift Universe
have formed up to 35\% of their stellar mass in this recent
episode - the mean value is 15\%. The star formation timescales
are large ($\sim$a few Gyrs), indicating that the SFR does not
decline significantly during the course of the burst.

14\% of the LIRG population host a Type 2 AGN. The AGN fraction
rises from 13.1\% for non-interacting objects to 17.6\% for the
interacting population and back down to 13.4\% for the
post-mergers, possibly supporting the hypothesis that interactions
trigger the onset of AGN. {\color{black}However, the numbers of
objects in each category are small and, while a trend does exist,
the low number statistics make it difficult to establish a robust
result.}

In agreement with previous studies, the distribution of bulge/disk
morphologies (derived from the SDSS \texttt{fracdev} parameter) in
the AGN hosts favours systems which have a significant bulge
component (\texttt{fracdev}$\gtrsim0.6$), in contrast to the
normal LIRGs which have a flat distribution in \texttt{fracdev}.

Comparison of the UV and optical colours of the AGN hosts to those
of the normal galaxies indicates that the AGN are consistently
redder than the normal population in both the UV and the optical
colours. We find that the AGN are redder in $(NUV-r)$, $(u-r)$ and
$(g-r)$ by $\sim1.2$, $\sim0.6$ and $\sim0.4$ mags respectively.
The systematic colour offsets persist even if we split the sample
into luminosity bins and/or explore different redshift ranges. The
distributions of $E_{B-V}$ values do not indicate a systematic
dust enhancement in the AGN hosts. Comparison of the derived SFH
parameters indicates that, while the underlying populations and
e-folding timescales of the star formation in AGN do not differ
from those in the normal galaxies, the ages of the RSF episodes in
the AGN are older (producing the redder colours), indicating that
we are simply observing the AGN hosts a \emph{longer} time after
the onset of star formation. The derived SFH parameters indicate
that the AGN typically appear $\sim0.5-0.7$ Gyrs after the onset
of star formation in their host galaxies.

The similarity between the SFH parameters of the AGN and the
normal galaxies indicates that the AGN do not have a measurable
impact on the star formation activity in their host galaxies. In
addition, we do not find evidence for a systematic deficit in the
cold gas content of the AGN compared to their normal counterparts,
which would be expected if feedback from the AGN was depleting the
cold gas reservoir in the host galaxies. We conclude that the AGN,
in this sample of objects, appear not to be exerting negative
feedback on the ISM by removing cold gas and reducing the star
formation activity.

While this issue needs to be investigated further, the lack of AGN
feedback in the LIRG population may have some intriguing
implications for the role of feedback envisaged in galaxy
formation models. For example, if we assume that low-redshift
LIRGs host similar conditions to star-forming objects at high
redshift, then our results imply that obscured AGN in such objects
may not significantly affect the star formation in their hosts
through e.g. mechanical `radio-mode' feedback
\citep[e.g.][]{Croton2006}. However, both observational
\citep[e.g.][]{Nesvadba2006} and theoretical
\citep[e.g.][]{Sijacki2007} evidence does exist for AGN feedback
in the high-redshift galaxy population and indirect evidence for
the radio-mode exists in low-redshift \emph{early-type} galaxies
\citep[e.g.][]{Best2005,Schawinski2007}. It is likely, therefore,
that conditions in low-redshift LIRGs do not mimic those in the
distant Universe ($z>1$). For example, gas densities at high
redshift are expected to be higher\footnote{The evolution of the
gas density shows a $(1+z)^{3/2}$ dependence with redshift
\citep{Rao1994}.} and galaxies are likely to be more compact,
plausibly allowing the AGN luminosity to couple more effectively
to the interstellar material. In addition, galaxies will tend to
be less massive, altering the proportional strength of the AGN
compared to the potential well of the host galaxy.

Perhaps more intriguing is the fact that (radio-mode) feedback
appears to operate in early-type galaxies at low redshift
\citep[][]{Schawinski2007}, while it is absent in the low-redshift
LIRGs (which are overwhelmingly late-type in terms of their
morphologies). Although we have demonstrated that LIRGs that host
AGN have prominent bulges, their eye-inspected morphologies show
that they are clearly late-type galaxies with a non-negligible
disk component. Mechanical feedback from AGN, in the conditions
that are prevalent at low redshift (e.g. gas densities, galaxy
masses etc), therefore seems to be most efficient in systems which
lack a significant disk component.

Finally, we have used the subsample of the LIRG population that
exhibits spheroidal or near-spheroidal morphology to explore the
properties of `mixed' major mergers in the local Universe.
Early-type systems in our LIRG sample - defined as spheroidal or
near-spheroidal objects that are plausible progenitors of
elliptical galaxies - have underlying populations that exhibit
(SSP-weighted) average ages of $\sim6.9$ Gyrs and form between 5
and 30\% of their stellar mass in the RSF episode, over time
periods between 0.3 and 4 Gyrs (lower limits since most of these
objects still show disturbed morphologies). This suggests that
these galaxies are the products of `mixed' major mergers, where at
least one of the progenitors has late-type morphology. Mergers
between two ellipticals would produce underlying populations that
are overwhelmingly old ($>10$ Gyrs old) and, since ellipticals are
typically gas-poor systems, a `dry' major merger (or equally a
minor merger, see e.g. Figure 1 in Kaviraj et al. 2007d) would not
supply enough gas to create the high LIRG-like SFRs in these
systems.

The local LIRG population offers a unique window into the vigorous
star formation that plausibly shaped much of the galaxy population
at high redshift. While these systems are rare at low redshift, an
improved understanding of stellar mass assembly in the Universe
will be greatly helped by studying the evolution of the LIRG
population to high redshift. The advent of deep optical surveys
offers an excellent opportunity to extend the analysis presented
in this paper by exploiting deep optical photometry to trace the
rest-frame UV of galaxy populations at intermediate and high
redshifts. In a forthcoming paper, we shall combine deep optical
surveys with Spitzer imaging of the Chandra Deep Field-South to
quantify the SFHs of distant LIRGs ($0.5<z<1.5$) and explore their
star formation properties. Combined with this study at low
redshift, these results will allow us better understand the
evolution of star-forming systems and the interplay between AGN
activity and star formation over the last 8-10 billion years of
cosmic time.


\nocite{Silva2004} \nocite{Neugebauer1984} \nocite{Taniguchi1997}
\nocite{Elbaz1999} \nocite{Veilleux1999} \nocite{Chakrabarti2007}
\nocite{Martin2005} \nocite{SDSSDR4} \nocite{Kaviraj2007a}
\nocite{Kaviraj2007b} \nocite{Kaviraj2007c} \nocite{Kaviraj2007d}
\nocite{Egami2004} \nocite{Bernardi2003b} \nocite{Gao2004a}
\nocite{Gao2004b} \nocite{Kaviraj2006a} \nocite{Kauffmann2003a}
\nocite{Kauffmann2003b} \nocite{Balwin1981}
\nocite{Brinchmann2004} \nocite{Burgarella2005a}
\nocite{Burgarella2005b} \nocite{Buat2005} \nocite{Strateva2001}
\nocite{Wyder2007} \nocite{Seibert2005} \nocite{Baldwin1981}


\section{Acknowledgements}
I acknowledge a Leverhulme Early-Career Fellowship, a Beecroft
Fellowship from the BIPAC institute at Oxford and a Research
Fellowship from Worcester College, Oxford. I thank Joseph Silk for
many useful discussions. I am grateful to Ted Wyder and Mark
Seibert for providing important information on pipeline processing
of publicly available GALEX GR3 data. Antonio Pipino, Sukyoung Yi,
Kevin Schawinski, Sadegh Khochfar, David Schiminovich and
Christian Wolf are thanked for constructive comments. I am
grateful to Christy Tremonti for useful discussions regarding the
extraction of spectroscopic parameters in the Garching SDSS
catalog that are used to derive the AGN classifications in this
paper.

Funding for the SDSS and SDSS-II has been provided by the Alfred
P. Sloan Foundation, the Participating Institutions, the National
Science Foundation, the U.S. Department of Energy, the National
Aeronautics and Space Administration, the Japanese Monbukagakusho,
the Max Planck Society, and the Higher Education Funding Council
for England. The SDSS Web Site is http://www.sdss.org/.

The SDSS is managed by the Astrophysical Research Consortium for
the Participating Institutions. The Participating Institutions are
the American Museum of Natural History, Astrophysical Institute
Potsdam, University of Basel, University of Cambridge, Case
Western Reserve University, University of Chicago, Drexel
University, Fermilab, the Institute for Advanced Study, the Japan
Participation Group, Johns Hopkins University, the Joint Institute
for Nuclear Astrophysics, the Kavli Institute for Particle
Astrophysics and Cosmology, the Korean Scientist Group, the
Chinese Academy of Sciences (LAMOST), Los Alamos National
Laboratory, the Max-Planck-Institute for Astronomy (MPIA), the
Max-Planck-Institute for Astrophysics (MPA), New Mexico State
University, Ohio State University, University of Pittsburgh,
University of Portsmouth, Princeton University, the United States
Naval Observatory, and the University of Washington.

GALEX (Galaxy Evolution Explorer) is a NASA Small Explorer,
launched in April 2003, developed in cooperation with the Centre
National d'Etudes Spatiales of France and the Korean Ministry of
Science and Technology.


\bibliographystyle{mn2e}
\bibliography{references}


\end{document}